\documentclass[12pt]{amsart}

\usepackage{amssymb}
\usepackage{hyperref}

\newtheorem{thm}{Theorem}[section]

\newtheorem{cor}[thm]{Corollary}
\newtheorem{lem}[thm]{Lemma}
\newtheorem{prop}[thm]{Proposition}

\newtheorem*{thm*}{Theorem}

\theoremstyle{definition}

\newtheorem*{defn*}{Definition}

\numberwithin{equation}{section}

\begin{document}

Published in: {\it International Mathematics Research Notices}\,
{\bf 2006} \, (2006), Article ID 61570, 37 pages

\title
 {On permanental polynomials of certain random matrices}

\author{Yan V Fyodorov}
\address{School of Mathematical Sciences,  University of Nottingham,  Nottingham NG7 2RD, United Kingdom}
\email{yan.fyodorov@nottingham.ac.uk}

\thanks{The present research is supported by EPSRC grant EP/C515056/1 "Random Matrices and Polynomials: a tool to understand
complexity"}

\thanks{Preprint AIM 2006-8, American Institute of Mathematics}

\maketitle

\begin{abstract}
The paper addresses the calculation of correlation functions of
permanental polynomials of matrices with random entries. By
exploiting a convenient contour integral representation of the
matrix permanent some explicit results are provided for several
random matrix ensembles. When compared with the corresponding
formulae for characteristic polynomials, our results show both
striking similarities and interesting differences. Based on these
findings, we conjecture the asymptotic forms of the density of
permanental roots in the complex plane for Gaussian ensembles as
well as for the Circular Unitary Ensemble of large matrix
dimension.

\end{abstract}

\section{Introduction}

The permanent of an $N\times N$ matrix ${\bf H}$ with entries
$H_{ij}\,\, (i,j=1,\ldots,N)$ is defined by
\begin{equation}\label{defperm}
Per({\bf H})=\sum_{\sigma}\prod_{i=1}^N H_{i\sigma(i)}
\end{equation}
where the sum is taken over all permutations $\sigma$ of
$\{1,\ldots,N\}$.
 In strong contrast to determinants,
computing permanents is $\#$-complete problem \cite{valiant}, and
a considerable effort was spent on developing various
approximation methods, see \cite{FRZ} and references therein.
Permanents have important combinatorial meaning, in particular the
permanents of $(0,1)$ matrices enumerate matchings in bipartite
graphs, see e.g.\cite{DiacGH}. Permanents also have applications
in physics of interacting Bose particles, see e.g. \cite{W}. A
concise introduction into properties of permanents can be found in
\cite{Krau1}, the standard reference is \cite{Minc}.

The permanental polynomial afforded by ${\bf H}$ is defined as the
permanent of the characteristic matrix, i.e.
\begin{equation}\label{perpol}
p\,(\mu)=Per\left(\mu{\bf 1}_N-{\bf
H}\right)=\mu^N-a_1\mu^{N-1}+a_2\mu^{N-2}-\ldots+(-1)^Na_N
\end{equation}
where ${\bf 1}_N$ stands for the $N\times N$ identity matrix.
 In particular, it is easy to show that $a_1=Tr{\bf H}$ and
$a_N=Per({\bf H})$. Although a permanental polynomial is not
preserved by similarity, it is preserved by permutational
similarity:
\begin{equation}\label{permut}
Per\left(\mu{\bf 1}_N-{\bf P}^{-1}{\bf H}{\bf
P}\right)=Per\left(\mu{\bf 1}_N-{\bf H}\right)
\end{equation}
for any $N\times N$ permutation matrices ${\bf P}$. Permanental
polynomials attracted some interest in graph theory and its
applications to chemistry, see e.g.\cite{Cash,Krau2,MRW}. In
particular, some general statements on location of the roots of
permanental polynomials (known as permanental roots) can be found
in \cite{Krau2,MRW}. Some curious series expansion of moments of
permanental polynomials were developed in \cite{chu}.

The goal of the present paper is to initiate research on
permanental polynomials of various classes of random matrices.
Although some questions concerning permanents of random matrices
were investigated earlier, see the books \cite{KB,Sachkov} as well
as more recent results in \cite{RW1,RW2}, the present author is
not aware of any related studies on permanental polynomials. The
final goal of the whole project should be finding various
statistical characteristics of the permanental roots,  the mean
density in the complex plane being the simplest, yet nontrivial
example. Formulated as such the problem seems to be quite
challenging, and no results are currently available.

To get some insight into the problem in the present paper we
address simpler, yet informative objects which we call the
correlation functions of permanental polynomials. The n-point
correlation function depends on $n$ complex parameters
$\mu_1,\ldots,\mu_n$ and is defined as the expectation value
\begin{equation}\label{corfun}
\left\langle Per\left(\mu_1{\bf 1}_N-{\bf
H}\right)Per\left(\mu_2{\bf 1}_N-{\bf H}\right)\ldots
Per\left(\mu_n{\bf 1}_N-{\bf H}\right)\right\rangle_{H},
\end{equation}
where the angular brackets stand for the averaging over the
distribution of the matrix entries of ${\bf H}$. Particular
classes of the distributions we are dealing with will be discussed
explicitly later on.

To this end it is appropriate to mention that in the last few
years it became clear that the general correlation functions of
characteristic polynomials $d(\mu)=\det{(\mu{\bf 1}_N-{\bf H})}$
of random matrices of various kinds are extremely informative
objects, with a rich mathematical structure and important
applications. One of the sources of interest in such objects
originated from attempts by Keating and Snaith\cite{KS1,KS2,KS3}
to get an analytical insight into statistical properties of the
Riemann zeta-function and related functions, whose zeroes are
believed to share statistically many features with the eigenvalues
of unitary (or Hermitian) random matrices. Brezin and Hikami
\cite{BH1,BH3} understood that the correlation functions of
products of characteristic polynomials for invariant ensembles of
random matrices can be represented as determinants made of
orthogonal polynomials generated by the random matrix measure, see
(\ref{bh2}) below. Fyodorov and Strahov generalized this result to
correlation functions involving both products and ratios of
characteristic polynomials\cite{FS,SF} (an elegant proof is by
Baik, Deift and Strahov \cite{BDS}, and important extensions to
other symmetry classes are due to Borodin and Strahov, \cite{BS}).
The latter objects are especially important for applications in
physics of quantum chaotic systems, see e.g. Andreev and
Simons\cite{AS}, and more recently in Quantum Chromodynamics , see
e.g. Verbaarschot and Splittorff \cite{SV}, Fyodorov and Akemann
\cite{FA,AF} and references therein. Characteristic polynomials of
non-Hermitian/non-unitary random matrices of several types were
studied most recently by Akemann and collaborators \cite{AV,AP}
and by Fyodorov and Khoruzhenko\cite{FK}.

 All these facts providing an additional strong
motivation for the research, a number of far-reaching
generalizations and extensions of the above-mentioned results were
obtained in recent years by various groups,
see\cite{BG,BOS,CFKRS,CFS,CFZ,HKO}. In addition, nice
combinatorial interpretations of moments of the characteristic
polynomials were recently revealed in
\cite{DiacGam,ForGam,Strahov}.

A major part of the recent progress in dealing with the
characteristic polynomials was essentially possible due to the
fact that those polynomials depended only on the eigenvalues
$\lambda_1,\ldots \lambda_N$ of random matrices, not on their
eigenvectors. The joint probability density (j.p.d.)
$\mathcal{P}(\lambda_1,\ldots,\lambda_N)$ of those eigenvalues is
well-known for the standard classes of random matrices.

As one of the most important examples, consider the space
$\mathcal{M}$ of $N\times N$ Hermitian matrices ${\bf H}
=(H_{ij})={\bf H}^*$, with the probability measure on
$\mathcal{M}$ being chosen according to
\begin{equation}\label{meas}
\mathcal{P}({\bf H})d{\bf H}=c\,e^{-NTr V({\bf
H})}\prod_{i=1}^NdH_{ii}\prod_{i<j}^N\,dH^{R}_{ij}\,dH^{I}_{ij}
\end{equation}
where $H_{ij}=H^{R}_{ij}+iH^{I}_{ij}$ in terms of its real and
imaginary parts, and $c$ stands for the corresponding
normalization constant. The function $V({\bf H})$ is known as the
{\it potential} and is usually assumed to be a polynomial in ${\bf
H}$ of even power with real coefficients, the coefficient in front
of the highest power being positive. Such potentials define the
so-called Unitary Ensembles \cite{De}, which is a distinguished
subset of distributions invariant with respect to unitary
conjugations: ${\bf H}\to {\bf U}{\bf H}{\bf U}^*$, with ${\bf U}$
standing for any $N\times N$ unitary matrix from the group $U(N)$.
Let us further use the spectral decomposition ${\bf H}={\bf
U}\Lambda{\bf U}^*$ of the Hermitian matrix ${\bf H}$ in terms of
the diagonal matrix $\Lambda=diag(\lambda_1,\ldots,\lambda_N)$ of
its (real) eigenvalues, and some ${\bf U}\in U(N)$. This
decomposition induces the corresponding decomposition of the
measure (see \cite{De,Fyo1,Me} for more detail):
\begin{equation}\label{factorize}
\mathcal{P}({\bf H})d{\bf H}=C_N\,e^{-NTr V({\bf
\Lambda})}\Delta^2(\Lambda) \prod_{i=1}^N\,d\lambda_{i}\,d\mu_N(U)
\end{equation}
where $d\mu_N(U)$ stands for the corresponding invariant (Haar's)
measure on $U(N)$ (normalized to unity),
$\Delta(\Lambda)=\prod_{i<j}(\lambda_i-\lambda_j)$ is the
so-called Vandermonde factor, and $C_N$ is the corresponding
normalization constant. Integrating out the $U$-variables
immediately gives j.p.d. of real eigenvalues in the form
\begin{equation}\label{eigenden}
\mathcal{P}(\lambda_1,\ldots,\lambda_N)\prod_{i=1}^N
d\lambda_{i}=C_N\ \prod_{i<j}^N (\lambda_i-\lambda_j)^2
\prod_{i=1}^N dw(\lambda_{i})
\end{equation}
where the measure $dw(\lambda)$ is related to the potential
$V(\lambda)$ as $dw(\lambda)=e^{-NV(\lambda)}d\lambda$. It is then
a straightforward exercise\cite{De,Fyo1} to show that the
expectation value $\langle d(\mu)\rangle_H$ of the characteristic
polynomial is simply given by
\begin{equation}\label{Heine}
\langle\det{(\mu{\bf 1}_N-{\bf H})}\rangle_H=\pi_N(\mu)
\end{equation}
where $\pi_k(x)$ stands for the $k-$th monic orthogonal polynomial
( $\pi_{k}(x)=x^{k}+$ lower powers), with the orthogonality being
understood with respect to the measure $dw(x)$ on the real axis:
\begin{equation}\label{ort}
\int \pi_j(x)\pi_k(x)dw(x) =c_jc_k\delta_{jk}
\end{equation}
In fact, the relation equivalent to (\ref{Heine}) was known in
classical theory of orthogonal polynomials (although, without any
reference to random matrices) for more than one hundred
years\cite{Szego}. It is its modern random matrix interpretation
that made evident a deep relation between the random
characteristic polynomials and orthogonal polynomials, and enabled
those far-reaching generalizations that were already mentioned
above. In particular, the simplest case of the Brezin-Hikami
result reads \cite{BH1}:
\begin{equation}\label{bh2}
\langle d(\mu_1)\,d(\mu_2)\rangle_{GUE}= \frac{1}{\mu_1-\mu_2}
\det{\left(\begin{array}{cc} \pi_N(\mu_1)& \pi_N(\mu_2)\\
\pi_{N+1}(\mu_1) & \pi_{N+1}(\mu_2)\end{array}\right)},
\end{equation}
\\

Naively one may expect that nothing similar is to be valid for
permanental polynomials, precisely because of the lack of
invariance under general similarity transformations. This fact
makes seemingly impractical presenting permanental polynomials in
terms of the matrix eigenvalues, and forces us to search for an
alternative technique of evaluating the random matrix averages
featuring in (\ref{corfun}), both over the eigenvalues and the
corresponding "angular" variables (=eigenvectors). To this end, we
start with revealing a possibility to represent permanents  as a
multiple contour integral, see Lemma (\ref{contint}). Such a
representation turned out to be very useful, and allowed us to
calculate explicitly the expectation value of the permanental
polynomial for a general Unitary Ensemble of Hermitian random
matrices. In fact, integrating out the "angular" variables can be
done explicitly with the help of a variant of the famous
HarishChandra-Itzykson-Zuber (HCIZ)formula\cite{HC,IZ}, and is
based essentially on interpreting that formula in terms of the
Schur function expansion method, see e.g.\cite{Or,BA}.
 It came as a somewhat
surprising result that the expectation value of the permanental
polynomial can be generally expressed as a one-fold integral of
the same monic orthogonal polynomials featuring in (\ref{Heine}),
see (\ref{principal}). In fact, we prove the following
\begin{thm}\label{thmglav}
The expected value of the permanental polynomial for random
matrices taken from a Unitary Ensemble characterized by a
potential $V(x)$ is given by
\begin{eqnarray}\label{principal}
\langle p\,(\mu)\rangle=a^{-1}_N\int e^{-N
V( \lambda)}(\mu-\lambda)^{2N-1}\pi_{N-1}(\lambda)\, d\lambda \\
\nonumber =a_N^{-1}\sum_{k=0}^Na_k\mu^k,\quad
a_k=\frac{(2N-1)!}{N!(N-1)!}\int e^{-N V(
\lambda)}\lambda^{2N-1-k}\pi_{N-1}(\lambda)\,d\lambda
\end{eqnarray}
where $\pi_{N}(\lambda)=\lambda^{N}+$ lower degrees is the $N-$th
monic polynomial orthogonal with respect to the measure
$dw(x)=e^{-NV(x)}dx $ on the real axis.
\end{thm}
For a particular case of the Gaussian Unitary Ensemble (GUE)
characterized by the measure $V(x)=x^2/2$ the integral in
(\ref{principal}) can be evaluated explicitly yielding , even more
surprisingly, the relation almost identical to (\ref{Heine}), with
the only simple change $\mu\to -i\mu$, see (\ref{GUE0}):
\begin{equation}\label{GUE00}
\langle p\,(\mu)\rangle_{GUE}=i^N\,\langle
d\,(-i\mu)\rangle_{GUE}, \quad d(\mu)=\det{(\mu{\bf 1}_N-H)}
\end{equation}

{\bf Note:} In this work we consider general  Hermitian matrices
with the eigenvalues $\lambda_i\in (-\infty,\infty)$. Sometimes,
however, one may wish to deal with positive definite Hermitian
matrices, and thus to confine the eigenvalues to the positive
semi-axis $\lambda_i\in [0,\infty)$ (e.g. the so-called Wishart
(or Laguerre) Ensemble, see \cite{For}, with $V(\lambda)=\lambda,
\,\, \lambda>0$). Further choices confining eigenvalues to an
interval of the real axis are also possible. Extensions of our
result to all those cases are self-evident. \\

The explicit evaluation of the lowest
 non-trivial correlation function of permanental polynomials
 turns out to be possible also for one more important class of
random matrices, those from the Circular Unitary Ensemble (CUE),
see Section \ref{Section5}. The corresponding matrices are general
unitary $N\times N$ and are considered to be uniformly distributed
on the group manifold according to the corresponding Haar's
measure. The calculation is based on a recent progress in
evaluating some integrals over $U(N)$ \cite{FK,SW}, and again has
the Schur function expansion in its heart.

Although in principle the Schur function expansion method can be
extended to obtain higher-order correlation functions, the actual
calculations for general invariant ensembles as well as for the
Circular Unitary Ensemble become cumbersome and the progress is
yet to be achieved. We therefore concentrate on two particular
cases where a progress can be achieved by a different method: the
Gaussian Unitary Ensemble of Hermitian matrices (Section
\ref{section4}) and the so-called Ginibre Ensemble (Section
\ref{section6}) of general complex matrices with independent,
identically distributed normal entries. Our starting expression is
the same contour integral representation of a permanent, but the
specific choice of the measure allows to perform the ensemble
averaging in a closed form without resorting to HCIZ formula.
Further progress is based on a method frequently employed in the
physics of disordered systems and known there as the
Hubbard-Stratonovich transformation, see e.g\cite{Fyo}. In this
way we managed to represent the general $n-$point correlation
function via $n-$fold integrals:
\begin{thm}\label{npointgue}
For any integer $n=1,2,\ldots$ the general $n-$ point correlation
function of permanental polynomials of GUE matrices has the
following integral representation:
\begin{equation}\label{maingau}
\left\langle \prod_{k=1}^n\,p(\mu_k)\right\rangle_{GUE}=\int
e^{-\frac{N}{2}Tr{\bf q}^2}\,\left[Per\left({\bf M}-{\bf
q}\right)\right]^N\,d{\bf q}
\end{equation}
In this expression ${\bf M}=diag(\mu_1,\ldots,\mu_n)$ is $n\times
n$ diagonal matrix and ${\bf q}$ is a general $n\times n$
Hermitian matrix.
\end{thm}
{\bf Note.} For $\mu_1=\mu_2=\ldots =\mu_N\equiv \mu$ the identity
(\ref{maingau}) can be written as
\begin{equation}\label{dualgau}
\int e^{-\frac{N}{2}Tr{\bf H}^2}\,\left[Per\left(\mu {\bf
1}_N-{\bf H}\right)\right]^n\,d{\bf H}=\int e^{-\frac{N}{2}Tr{\bf
q}^2}\,\left[Per\left(\mu{\bf 1}_n-{\bf q}\right)\right]^N\,d{\bf
q}
\end{equation}
Similar identities are known to hold for moments of GUE
characteristic polynomials, and are frequently called "the duality
relations", see e.g. \cite{FS,For} and references therein. In that
context they were commonly thought to be intimately connected to
the underlying integrable structures (Toda lattice hierarchies,
see e.g. \cite{For}). The fact that they emerge in the present
context may indicate some hidden integrability lurking behind the
problem of permanental polynomials.

 The formula (\ref{maingau}) is already a substantial simplification, especially if we
have in mind random matrices of large size $N$, so that $n$ can be
small in comparison with the size of the matrix. Nevertheless, to
get a closed form result for any $n$ and $N$ amounts to evaluating
those integrals and remains an outstanding problem. However, for
the two-point correlation functions $n=2$ the integral can be
evaluated explicitly, and the result is expressed in a very
attractive "determinantal" form which is actually again almost
identical to (\ref{bh2}) for characteristic polynomials. Namely,
we found that
\begin{equation}\label{relgue}
\langle p(\mu_1)\,p(\mu_2)\rangle_{GUE}=\langle
d(-i\mu_1)\,d(i\mu_2)\rangle_{GUE}\,.
\end{equation}
We also show that a formula similar to (\ref{maingau}) can be
derived by essentially the same method for the Ginibre Ensemble,
see (\ref{maingin}). Finally, the same method works also for the
ensemble of real symmetric matrices with the Gaussian probability
measure (the so-called Gaussian Orthogonal Ensemble, GOE), for
which we again obtain explicit expressions for the one- and
two-point correlation functions of permanental polynomials, see
Section\ref{section7}. Close similarities with the corresponding
results obtained earlier\cite{BH3} for characteristic polynomials
of GOE matrices are again apparent. In particular, both the
relations (\ref{GUE00}) and (\ref{relgue}) remain valid.

The last (but by far not the least) point to be mentioned is that
one is usually interested in the asymptotic, large-$N$ behaviour
of random matrix characteristics. In particular, for GUE
characteristic polynomials asymptotic behaviour is very different
for $\mu\in (-2,2)$ (the so-called "bulk scaling" regime), for
$|\mu|=2+O(N^{-2/3})$ ("soft edge scaling") and for all other
complex values of $\mu$ \cite{De}. This is naturally related to
the fact that all $N$ eigenvalues $\lambda_i$ of GUE matrices,
being real, concentrate in the limit $N\to \infty$ in the interval
$\lambda\in[-2,2]$ with probability tending to one. The
corresponding limiting density of eigenvalues is given by the
famous Wigner semicircular law:
$\rho_{\infty}(\lambda)=\frac{1}{2\pi}\sqrt{4-\lambda^2}$.
Moreover, within that interval those eigenvalues are non-trivially
correlated on the distances $|\lambda_i-\lambda_j|$ of the order
of $O(1/N)$. The latter fact is reflected, in particular, in the
existence of the following limit,\cite{BH1}, see also \cite{Fyo1}:
\begin{equation}\label{guecorr}
\lim_{N\to \infty}\frac{\langle
d(\mu_1)\,d(\mu_2)\rangle_{GUE}}{\langle
d^2(\mu_1)\rangle_{GUE}}=K_{\infty}[N\rho_{\infty}(\mu)(\mu_1-\mu_2)],\quad
K_{\infty}[x]=\frac{\sin{x}}{x}
\end{equation}
whenever $\mu_1,\mu_2\in (-2,2),\, \lim_{N\to \infty}
N(\mu_1-\mu_2)<\infty$. Moreover, the limiting expression
(frequently called the "Dyson kernel") is not restricted by GUE,
but is valid for a very broad class of Unitary Invariant Ensembles
of random Hermitian matrices ("universality").

Adopting those facts to the case of GUE permanental polynomials by
the relation (\ref{GUE00}) we see a special role of the interval
$\mu\in [-2i,2i]$ of the imaginary axis. Combining (\ref{relgue})
and (\ref{guecorr}) we also see that the values of permanental
polynomials close to the points $i\mu$ and $-i\mu$ are
non-trivially correlated in the limit of large $N$, as long as $\,
\mu\in (2,2)$. The latter fact could probably be related to the
property of the permanental roots to come in complex conjugate
pairs (as is evident from coefficients of the permanental
polynomials of Hermitian matrices being real). All this facts
taken together make it rather plausible to expect permanental
roots of large GUE matrices to accumulate asymptotically in the
vicinity of the interval $[-2i,2i]$ of the imaginary axis. Below
we provide an argument in favour of the validity of the
following\\
\noindent {\bf Conjecture}: {\it The normalized asymptotic
limiting density of the permanental roots of large GUE matrices in
the complex plane $z=x+iy$ is vanishing outside the interval
$[-2i,2i]$ of the imaginary axis. Inside that interval
it is given by the Wigner semicircular law.} \\

Our argument goes as follows. For $z=x+iy$ define the function
\begin{equation}\label{phi}
\Phi(x,y)=\lim_{N\to \infty}\frac{1}{N}\langle
\ln{|p(z)|^2}\rangle_{GUE}
\end{equation}
According to the standard potential theory \cite{Kellog} the
expectation value of the limiting density $\rho(x,y)$ of
permanental roots in the complex plane is given by
$\langle\rho(x,y)\rangle_{GUE}=\frac{1}{4\pi}\Delta \Phi(x,y)$,
with $\Delta$ standing for the Laplacian in variables $x$ and $y$
(understood in the sense of distributions). Unfortunately,
performing the ensemble average in (\ref{phi}) is an outstanding
problem. Assume however that the operations of taking the ensemble
average and taking the logarithm commute in the limit $N\to
\infty$, i.e.
\begin{equation}\label{commute}
\lim_{N\to \infty}\frac{1}{N}\langle \ln{|p(z)|^2}\rangle_{GUE}=
\lim_{N\to \infty}\frac{1}{N}\ln{\langle |p(z)|^2\rangle_{GUE}}
\end{equation}
Such commutativity is indeed known to occur, for example, in the
case of characteristic polynomials of GUE matrices\cite{Berezin},
and one may hope it to hold for permanental polynomials as well.
It would also imply the self-averaging property: the asymptotic
root density is given by its expectation value. Proving relation
(\ref{commute}) for GUE permanental polynomials remains an
outstanding problem. If however such a relation holds, the
function $\Phi(x,y)$ can be found from
(\ref{commute}),(\ref{relgue}) and (\ref{bh2}) by exploiting the
known Plancherel-Rotach asymptotics for the Hermite polynomials
$\pi_N(\mu)$, see e.g. Eq.(7.93) of \cite{De}. A simple
calculation yields for $x\ne 0$
\begin{eqnarray}\label{phi1}
&&\Phi(x,y)=\Psi(y-ix)+\Psi(y+ix)\\&&
\Psi(q)=\frac{1}{8}(q-\sqrt{q^2-4})^2-\ln{(q-\sqrt{q^2-4})}+const
\end{eqnarray}
For $x=Im\,q\ne 0$ the function $\Psi(q)$ is obviously analytic in
the plane of complex variable $q=y+ix$, hence the function
$\Phi(x,y)$ is harmonic resulting in vanishing root density
$\rho(x,y)=0$. Thus, the only non-vanishing root density is
possible for $x=0$, i.e. along the imaginary axis in the original
complex plane $z=x+iy$. To calculate this density from $\Phi(x,y)$
given in (\ref{phi1}) one again follows the standard potential
theory and evaluates the jump of the normal derivative:
$\frac{\partial}{\partial x}\Phi(x,y)|_{x\to
0^+}-\frac{\partial}{\partial x}\Phi(x,y)|_{x\to 0^-}$ across the
line $x=0$. It immediately yields the Wigner semicircular law in
the interval $[-2i,2i]$ of the imaginary axis, in agreement with
the proposed conjecture. Similar, but slightly more technically
involved calculations suggest validity of the semicircular density
profile at the interval $[-2\sqrt{2}\,i,2\sqrt{2}\, i]$ of the
imaginary axis for the permanental roots of Gaussian Orthogonal
Ensemble of real symmetric random matrices.

The relation (\ref{commute}) is known to hold also for
characteristic polynomials of Ginibre Ensemble and Circular
Unitary Ensemble (CUE) matrices, see \cite{FK} and the discussion
at the end of Sections \ref{Section5} and \ref{section6} .
Assuming its validity for permanental polynomials of those
ensembles one can conjecture the ensuing asymptotic densities of
the permanental roots for those ensembles inside the unit circle
$x^2+y^2\le 1$ in the complex plane. They are given by:
\begin{eqnarray}\label{dencomp1}
&&\rho(x,y)=\frac{2}{\pi}\frac{1}{(1+x^2+y^2)^2},\quad
\mbox{CUE}\\
&&\rho(x,y)=\frac{1}{\pi},\quad \mbox{Ginibre
Ensemble}\label{dencomp2}
\end{eqnarray}
and vanishing density outside the unit circle. Whereas for Ginibre
ensemble the conjectured density of permanental roots
(\ref{dencomp2}) is the same as that for the characteristic roots,
the density profile (\ref{dencomp1}) is obviously radically
different from the CUE eigenvalue density concentrated on the unit
circle. Apart from investigating the validity of the above
conjectures, any result on correlation patterns typical for
permanental roots of large random matrices of various types are
presently lacking. Finally, the question whether an accumulation
of permanental roots in the vicinity of the imaginary axis is
generic for more general ensembles of Hermitian random matrices is
very intriguing and clearly deserve to be a subject of a separate
study.

{\bf Acknowledgements}. The author is grateful to Boris
Khoruzhenko for numerous stimulating discussions at various stages
of this work, and in particular for providing access to the
references \cite{KB} and \cite{Berezin}. Alexander Gamburd is
acknowledged for discussion of results at early stages of the
research, and Ilya Krasikov for a few useful remarks and
encouragement. Although first results in this direction were
obtained in 2004, the paper would be hardly completed, had it been
not for the author's participation in January 2006 at the workshop
"Random Analytic Functions" at American Institute of Mathematics,
Palo Alto. The author is very grateful to the organizers,
coordinators and participants of the workshop for numerous
informative discussions and inspiring atmosphere, in particular to
Pavel Blecher, Brian Conrey, David Farmer and especially to
Stanislav Molchanov for his remarks on the conjecture concerning
the asymptotic density of permanental roots.

\section{Permanents as contour integrals}\label{section 2}
Let ${\bf z}=(z_1,\ldots,z_N)^T$ stand for $N-$ component complex
(column) vector, and similarly define a (row) vector ${\bf
\xi}^*=(\overline{\xi}_1,\ldots,\overline{\xi}_N)$, with the bar
standing for the complex conjugation, and the star for Hermitian
conjugation. Further denote ${\bf z}\otimes \xi^*$ the $N\times N$
matrix with entries $ z_i\overline{\xi}_j$, $\, i,j=1,\ldots,N$,
and let $(\xi^*{\bf z})$ stand for the scalar product
$\sum_{i=1}^N \overline{\xi}_iz_i$, and ${\bf \xi}^*{\bf F}{\bf
z}$ for the bilinear form $\sum_{ij}^N F_{ij}\overline{\xi}_iz_j$.

\begin{lem}\label{contint}
The permanent of an arbitrary $N\times N$ matrix ${\bf
F}=\{F_{ij}\}$ can be expressed through the contour integral as
\begin{eqnarray}\label{main}
&& \,\, Per\, {\bf F}=\frac{1}{(2\pi)^{2N}}\oint_{|z_1|=1} \ldots
\oint_{|z_N|=1}\oint_{|\xi_1|=1} \ldots \oint_{|\xi_N|=1}
\\\nonumber &\times& \exp{Tr\left[{\bf F}\,{\bf z}\otimes
{\xi}^*\right]}
 \prod_{k=1}^N\frac{dz_k}{z_k^2}\,\,\frac{d\overline{\xi}_k}{\overline{\xi}_k^2}
\end{eqnarray}
\end{lem}

\begin{proof}
Obviously, $Tr\left[{\bf F}\,{\bf z}\otimes  {\xi}^*\right]={\bf
\xi}^*{\bf F}{\bf z}$ for any matrix ${\bf F}$. Denote ${\bf
q}={\bf F}{\bf z}$, so that $\exp{\left({\bf \xi}^*{\bf F}{\bf
z}\right)}=\prod_{i=1}^N \exp{\overline{\xi}_iq_i}$. Use
$\frac{1}{2\pi i} \oint_{|\xi|=1}
\frac{d\overline{\xi}}{\overline{\xi}^2}\exp{\overline{\xi}q}=-q$
to see that the right-hand side of (\ref{main}) is given by
\begin{eqnarray}\label{main1}
\,\, (-i)^N \frac{1}{(2\pi)^{N}}\oint_{|z_1|=1}\ldots
\oint_{|z_N|=1}
\prod_{i=1}^N\sum_{j=1}^N {F}_{ij} z_j\, \prod_{k=1}^N \frac{dz_k}{z_k^2} \\
\nonumber = \frac{1}{(2\pi i)^N}\sum_{j_1=1,\ldots,j_N=1}^N\, {
F}_{1j_1}\ldots { F}_{Nj_N}\oint_{|z_1|=1} \ldots \oint_{|z_N|=1}
\,\, z_{j_1}z_{j_2}\ldots z_{j_N}\, \prod_{k=1}^N
\frac{dz_k}{z_k^2}
\end{eqnarray}
The last integral is obviously non-vanishing only as long as
$j_1\ne j_2 \ne \ldots \ne j_N$, and the required relation
immediately follows.
\end{proof}

Our way of representing permanents as contour integrals is of
central importance for the rest of the paper. Despite its apparent
simplicity,  we were not able to trace such a formula in the
available literature on permanents. In the Appendix we demonstrate
how the formula (\ref{main}) generates a known integral
representation of the permanent of a positive definite matrix as a
multivariate Gaussian integral used earlier in the physical
applications \cite{W}.

In what follows we will systematically use the short-hand notation
\begin{eqnarray}\label{shorthand}
 && \oint \left[\ldots\right]
\mathcal{D}_N({\bf z},\xi^*)\\ \nonumber &\equiv&
\frac{1}{(2\pi)^{2N}}\oint_{|z_1|=1} \ldots \oint_{|z_N|=1}
\oint_{|\xi_1|=1} \ldots \oint_{|\xi_N|=1} \left[\ldots\right]
\prod_{k=1}^N\frac{dz_k}{z_k^2}\prod_{k=1}^N
\frac{d\overline{\xi}_k}{\overline{\xi}_k^2}
\end{eqnarray}

In particular, we will make use of the following
\begin{lem}
Let $f(z)$ be an entire function of the complex variable $z$
represented by its convergent power series
$f(z)=\sum_{n=0}^{\infty}f_nz^n$. Then:
\begin{equation}\label{taylor}
I_N(f)=\oint  f(\xi^*{\bf z})\,\mathcal{D}_N({\bf z},\xi^*)=f_N
\end{equation}
\end{lem}
\begin{proof} Use $z_k=e^{i\theta_k},\,\,
\xi_k=e^{i\psi_k},\,\,k=1,\ldots,N$, where $0\le
\theta_k,\psi_k<2\pi$, so that
\[
I_N(f)=\left(\frac{i}{2\pi}\right)^{2N}\int_0^{2\pi}\int_0^{2\pi}
e^{-i\sum_k(\theta_k-\psi_k)}
f\left(\sum_ke^{i(\theta_k-\psi_k)}\right)\,\,\prod_k
d\theta_k\prod_k d\psi_k\,\,
\]
Shifting $\theta_k\to \alpha_k=\theta_k-\psi_k$, integrating out
$\psi_k$ and further reinstating the unimodular complex variable
$z=e^{i\alpha_k}$ we see that
\[
I_N(f)=\left(\frac{i}{2\pi}\right)^{N}\oint_{|z_1|=1} \ldots
\oint_{|z_N|=1} f\left(\sum_{k} z_k\right)\frac{dz_1}{z_1^2}\ldots
\frac{dz_N}{z_N^2}
\]
The integrations can be easily performed using the power series
expansion, and the result follows.
\end{proof}

\section{Mean value of the permanental polynomials
for unitary invariant ensembles of Hermitian
matrices}\label{section 3}

The main goal of the present section is to verify the formula
(\ref{principal}). We need the following
\begin{lem}\label{lemmaIZ}
  Let $\Gamma$ be any $N\times N$ matrix such that it
has precisely one non-zero eigenvalue $\gamma$, the rest $N-1$
eigenvalues being equal to zero. Then the following identity
holds:
\begin{eqnarray}\label{IZ}
I(\beta;\Lambda,\Gamma)&=&\int_{U(N)}\,\exp{\beta Tr\left[{\bf
U}\Lambda {\bf U}^* \,\Gamma\right]}\,\,d\mu_N(U)\\
\label{rankone} &=&
\sum_{n=0}^{\infty}\frac{(N-1)!}{(N+n-1)!}(\beta\gamma)^{n}
h_n(\lambda_1,\ldots,\lambda_N)
\\ &=&\frac{(N-1)!}{(\beta\gamma)^{N-1}}\sum_{i=1}^N
\frac{e^{\beta\gamma\lambda_i}}{\prod_{j\ne
i}(\lambda_j-\lambda_i)}\label{rankone1}
\end{eqnarray}
where
\begin{equation}\label{sym}
h_n(\lambda_1,\ldots,\lambda_N)=\sum_{1\le i_1\le i_2\ldots \le
i_n\le N} \lambda_{i_1}\lambda_{i_2}\ldots \lambda_{i_n}
\end{equation}
are the complete symmetric functions.
\end{lem}

{\bf Remark:} The integral in the right hand side of (\ref{IZ}) is
the well-known HarishChandra-Itzykson-Zuber (HCIZ)
integral\cite{HC,IZ}. In the case of a general Hermitian matrix
$\Gamma$ with all distinct non-zero real eigenvalues
$\gamma_1,\ldots,\gamma_N$ the result was found by Itzykson and
Zuber \cite{IZ} in the form
\begin{equation}\label{IZHCfull}
I(\beta;\Lambda,\Gamma)=\beta^{-N(N-1)/2}\left(\prod_{n=0}^{N-1}n!\right)\frac{\det{e^{\beta\lambda_i\gamma_j}}}{\Delta(\Lambda)\Delta(\Gamma)}.
\end{equation}
In our case it is most convenient to exploit that fact that one
can recover HCIZ expression by the Schur function expansion
method\cite{Or} (also known as the character expansion, see
\cite{BA}). In particular, the method does not require matrices
$\Lambda$ and $\Gamma$ to be Hermitian, and we will indeed
consider them to be general complex.

\begin{proof}

Let us shortly remind the definition and basic properties of the
Schur functions (see the book \cite{Ma} for more detail). A
partition $r=(n_1,n_2,\ldots,n_N)$ is a finite sequence of
integers, called "parts", satisfying $(n_1\ge n_2\ge \ldots\ge
n_N)$, which is characterized by its weight $|r|=\sum_j n_j$ and
its length $l(r)$ equal to the number of non-zero parts. With any
partition of the length $l(r)\le N$ one associates the Schur
function
\begin{equation}\label{Schur}
s_r(x_1,x_2,\ldots,x_N)=\frac{\det{\left(x_i^{n_j+N-j}\right)_1^N}}
{\det{\left( x_i^{N-j}\right)_{1}^N}}
\end{equation}
which is a symmetric polynomial in $x_1,x_2,\ldots,x_N$
homogeneous of degree $|r|$. By convention,
$s_r(x_1,\ldots,x_N)=0$ for $l(r)> N$.

For the simplest case of partitions of length one $r=(n)$ (i.e.
$n_1=n,\,n_2=\ldots=n_N=0$) the Schur functions coincide with the
complete symmetric functions:
\begin{equation}\label{l=1}
s_{(n)}(x_1,\ldots,x_N)=h_n(x_1,\ldots,x_N) \end{equation}
 defined in (\ref{sym}). Using definition (\ref{Schur}), expanding the
determinant in the numerator in entries of the first column and
remembering \[\det{\left( x_i^{N-j}\right)}_{1}^N=\prod_{i<
j}(x_i-x_j)\] we easily get a useful identity:
\begin{equation}\label{symfun1}
(-1)^{N-1}\,\sum_{i=1}^N\frac{x_i^{n+N-1}}{\prod_{j\ne
i}(x_j-x_i)}=\left\{\begin{array}{c}0,\quad
n=-(N-1),-(N-2),\ldots,-1\\  h_n(x_1,\ldots,x_N),\quad
n=0,1,2,\ldots  \end{array} \right.
\end{equation}

For any $N\times N$ matrix ${\bf M}\in Gl(N)$ one defines Schur
functions of the matrix argument, $s_r({\bf M})$, as
\begin{equation}\label{matrixschur}
s_r({\bf M})=s_r(m_1,\ldots,m_N)
\end{equation}
where $m_k$ are eigenvalues of the matrix ${\bf M}$. The Schur
functions of matrix argument are the characters of irreducible
representations of the general linear group (and its unitary
subgroup), and as a consequence have the important property of
orthogonality. If $r_1$ and $r_2$ are two partitions, and ${\bf
M}_1,{\bf M}_2$ are two $N\times N$ matrices, then (see p.445 of
\cite{Ma})
\begin{equation}\label{ort1}
\int_{U(N)}\, s_{r_1}({\bf UM}_1) \overline{s_{r_2}({\bf
UM}_2)}\,\,d\mu_N(U)=\delta_{r_1,r_2}\frac{s_{r_1}({\bf M}_1{\bf
M}_2^* )}{s_{r_1}({\bf 1}_N)}
\end{equation}
and
\begin{equation}\label{ort2}
\int_{U(N)}\, s_{r} ({\bf UM}_1{\bf U^*M}_2)
\,\,d\mu_N(U)=\frac{s_{r}({\bf M}_1)s_r({\bf M}_2)}{s_{r}({\bf
1}_N)}
\end{equation}
Using all these facts, one can calculate the integral in
(\ref{IZ}) by expanding the integrand in a series with respect to
Schur functions, and performing the integration over the unitary
group using (\ref{ort2}). After working out the coefficients of
expansion the result is expressed as a sum over
partitions\cite{BA,Or}
\begin{equation}\label{IZ1}
I(\beta;\Lambda,\Gamma)=\sum_r\beta^{n_1+\ldots+n_N} \prod_{i=1}^N
\frac{(N-i)!}{(N+n_i-i)!}s_r(\Lambda)s_r(\Gamma)
\end{equation}
for general matrices $\Lambda$ and $\Gamma$. In our particular
case $\gamma_2=\ldots=\gamma_N=0$, and therefore
$s_r(\Gamma)=s_r(\gamma_1)=\gamma_1^n$ for $r=(n)$, i.e for
partitions of the length $l(r)= 1$, and $s_r(\Gamma)=0$ for
$l(r)>1$. Using (\ref{l=1}) the expression (\ref{IZ1}) immediately
takes the form of (\ref{rankone}), and the equivalence to
(\ref{rankone1}) follows after exploiting the identity
(\ref{symfun1}).
\end{proof}

Now we are in position to prove {\bf Theorem (\ref{thmglav})}.
\begin{proof}
Our starting point is  the integral representation (\ref{main})
according to which the permanental polynomial
$p\,(\mu)=Per(\mu{\bf 1}_N-{\bf H})$ is given by
 \begin{equation}\label{start}
 p\,(\mu)=\oint \exp{\mu(\xi^*{\bf z})}\exp{-Tr\left[{\bf H}\,{\bf
z}\otimes {\xi}^*\right]}\,\mathcal{D}_N({\bf z},\xi^*)
 \end{equation}

According to (\ref{factorize}),(\ref{eigenden}) the ensemble
average of this polynomial then amounts to evaluating the integral
\begin{equation}\label{st1}
\langle p\,(\mu)\rangle= C_N\int_{-\infty}^{\infty}  e^{-N\,Tr
V({\bf \Lambda})}\Delta^2(\Lambda)\,\prod_{i=1}^Nd\lambda_{i}
\oint
 e^{\mu(\xi^*{\bf z})}J_N({\bf z},\xi^*)\, \mathcal{D}_N({\bf z},\xi^*)
\end{equation}
where
\begin{equation}\label{un}
J_N({\bf z},\xi^*)=\int_{U(N)} \exp{-Tr\left[{\bf U}\Lambda{\bf
U}^* \,{\bf z}\otimes \xi^*\right]}\,\,d\mu_N(U)
\end{equation}

For performing the latter integration over the unitary group we
can use the Lemma (\ref{lemmaIZ}) due to the fact that the
(non-Hermitian) matrix ${\bf z}\otimes \xi^*$ has precisely one
non-trivial eigenvalue equal to the scalar product $( \xi^*{\bf
z})$. We find

\begin{equation}\label{un1}
J_N({\bf z},\xi^*)=
\sum_{n=0}^{\infty}\frac{(N-1)!}{(N+n-1)!}(-1)^n(\xi^*{\bf z})^{n}
h_n(\lambda_1,\ldots,\lambda_N)
\end{equation}

Substituting this back to (\ref{st1}) we at the next step should
evaluate the contour integral according to (\ref{taylor}). For
doing this we need to find $N-$th term in the Taylor expansion of
$e^{\mu(\xi^*{\bf z})}J_N({\bf z},\xi)$ in powers of $(\xi^*{\bf
z})$. This is obviously given by
\begin{equation}\label{summa}
\sum_{l=0}^N\frac{\mu^{N-l}}{(N-l)!}
\frac{(N-1)!}{(N+l-1)!}(-1)^lh_l(\lambda_1,\ldots,\lambda_N)
\end{equation}
\[=\frac{(N-1)!}{(2N-1)!}\sum_{i=1}^N\frac{1}{\prod_{j\ne
1}(\lambda_j-\lambda_1)}\sum_{l=-(N-1)}^N\frac{(2N-1)!}{(N-l)!(N+l-1)!}(-1)^{l+N-1}\mu^{N-l}\lambda_i^{l+N-1}
\]
where we exploited the identity (\ref{symfun1}), which allowed us
to replace the lower limit $l=0$ in the second sum in
(\ref{summa}) with $l=-(N-1)$. Now the sum over $l$ gives simply
$(\mu-\lambda_i)^{2N-1}$, and we arrive at the result of the
contour integration in the form
\begin{equation}\label{st2}
\oint e^{\mu(\xi^*{\bf z})}J_N({\bf z},\xi)\, \mathcal{D}_N({\bf
z},\xi^*)=\frac{(N-1)!}{(2N-1)!}\sum_{i=1}^N
\frac{(\mu-\lambda_i)^{2N-1}}{\prod_{j\ne i}(\lambda_j-\lambda_i)}
\end{equation}

Next step is to substitute (\ref{st2}) to (\ref{st1}), and to use
the symmetry of the integrand with respect to a permutation of
variables $\lambda_1,\ldots,\lambda_N$. This gives
\begin{equation}\nonumber
\langle p\,(\mu)\rangle= \tilde{C}_N\int e^{-N\sum_{k=1}^N V(
\lambda_k)}\frac{(\mu-\lambda_1)^{2N-1}}{\prod_{j\ne
1}(\lambda_j-\lambda_1)} \prod_{j\ne
i}(\lambda_j-\lambda_i)^2\,\prod_{i=1}^Nd\lambda_{i}
\end{equation}
\begin{eqnarray}\label{st4}
=\tilde{C}_N\int e^{-N V( \lambda_1)}(\mu-\lambda_1)^{2N-1}
d\lambda_1\\ \nonumber \times \int e^{-N\sum_{k=2}^N V(
\lambda_k)} \prod_{j=2}^N(\lambda_1-\lambda_j) \prod_{2\le j\ne
i\le N }(\lambda_j-\lambda_i)^2\,\prod_{i=2}^Nd\lambda_{i}
\end{eqnarray}
where the particular value  of the proportionality factor
$\tilde{C}_N$ is not needed for our purposes, and can be
established whenever necessary ( see below).

 According to
the classical theory of orthogonal polynomials
\cite{Szego,De,For}, the integral in the last line of (\ref{st4})
is proportional (see (\ref{Heine})) to the monic orthogonal
polynomial: $\pi_{N-1}(\lambda_1)=\lambda_1^{N-1}+$ lower degrees,
defined in (\ref{ort}). The overall proportionality constant can
be fixed by the obvious condition $\langle p\,(\mu)\rangle=\mu^N+$
lower order terms, and we immediately arrive at the statement of
the Theorem (\ref{thmglav}).
\end{proof}

\begin{cor}
For the Gaussian Unitary Ensemble (GUE) corresponding to the
choice $V(x)=x^2/2$ the relation (\ref{principal}) takes the form
\begin{equation}\label{GUE0}
\langle p\,(\mu)\rangle_{GUE}=i^N\,\pi_{N}(-i\mu),\quad
\end{equation}
\end{cor}
\begin{proof}
The orthogonal polynomials are obviously Hermite polynomials. The
corresponding monic polynomials $\pi_k(x)$ have the following
convenient integral representation, see e.g p.53 of \cite{Fyo1}:
\begin{equation}\label{intrep}
\pi_{k}(x)=(-i)^k\sqrt{\frac{N}{2\pi}}e^{Nx^2/2}\int_{-\infty}^{\infty}
q^{k}e^{-N\frac{q^2}{2}+iNxq}\,dq
\end{equation}
Substitute (\ref{intrep}) to (\ref{principal}) and change
$\mu-\lambda\to x$, obtaining
\begin{eqnarray}\nonumber\,\langle
\,p\,(\mu)\rangle_{GUE}=-\frac{i^{N-1}}{a_N}\sqrt{\frac{N}{2\pi}}\int_{-\infty}^{\infty}x^{2N-1}\,dx
\int_{-\infty}^{\infty}
q^{N-1}e^{-N\frac{q^2}{2}+iN(\mu-x)q}\,dq\\
\nonumber =\frac{(-1)^N}{a_NN^{N-1}}\sqrt{\frac{N}{2\pi}}
\int_{-\infty}^{\infty}x^{2N-1}
\frac{d^{N-1}}{dx^{N-1}}\left(\int_{-\infty}^{\infty}
e^{-N\frac{q^2}{2}+iN(\mu-x)q}\,dq\right)\,dx\\ \nonumber
=\frac{(-1)^N}{a_N N^{N-1}} \int_{-\infty}^{\infty}x^{2N-1}
\frac{d^{N-1}}{dx^{N-1}} \left(
e^{-\frac{N}{2}(\mu-x)^2}\right)\,dx\\
\label{g} = -\frac{1}{a_N}\frac{(2N-1)!}{N^N (N-1)!}\,
e^{-\frac{N}{2}\mu^2}
\int_{-\infty}^{\infty}x^Ne^{-\frac{N}{2}x^2+N\mu x}\,dx
\end{eqnarray}
The same method can be used to find the normalization constant
from (\ref{principal}) and (\ref{intrep}):
\begin{equation}\label{normal}
a_N=\frac{1}{N^{N}}\frac{(2N-1)!}{(N-1)!}\sqrt{\frac{2\pi}{N}}
\end{equation}

Comparing (\ref{g},\ref{normal}) with (\ref{intrep}) immediately
verifies (\ref{GUE0}).
\end{proof}
In the next section we will be able to reproduce the above result
for GUE case by a different method, which is essentially
originating from applications of random Gaussian matrices in
physics of disordered systems, see \cite{Fyo} and references
therein. In this way we will be able also to get some insight into
higher order correlation functions of GUE permanental polynomials.

\section{Correlation functions for permanental polynomials of GUE
matrices}\label{section4}
 The joint probability density of matrix
entries for GUE $N\times N$ Hermitian matrices ${\bf H}$ is given
by
\begin{equation}\label{GUEmeas}
\mathcal{P}_{GUE}({\bf H})d{\bf H}=C_{GUE} \prod_{i=1}^N
e^{-\frac{N}{2}H_{ii}^2}\prod_{i<j}^N e^{-N
\overline{H}_{ij}H_{ij}}
\prod_{i=1}^NdH_{ii}\prod_{i<j}^N\,dH^{R}_{ij}dH^{I}_{ij}
\end{equation}
where $C_{GUE}=2^{-\frac{N}{2}}\left(N/\pi\right)^{N^2/2}$ is the
corresponding normalization constant.

\begin{lem}
 The identity
\begin{equation}\label{gueav}
\left\langle e^{-Tr {\bf H A}}
\right\rangle_{GUE}=e^{\frac{1}{2N}Tr{\bf A}^2}
\end{equation}
holds for a general $N\times N$ matrix ${\bf A}$ with complex
entries.
\end{lem}
\begin{proof}
The integral in the left hand side factorizes into a product of
simple Gaussian integrals of the sorts
\[
\int e^{-\frac{N}{2}H_{ii}^2-H_{ii}A_{ii}} dH_{ii}=
\sqrt{\frac{2\pi}{N}}\,\exp{\left[\frac{1}{2N}A_{ii}^2\right]}
\]
and
\[
\int e^{-N
\overline{H}_{ij}H_{ij}-H_{ij}A_{ji}-\overline{H}_{ij}A_{ij}}
dH^{(R)}_{ij}dH^{(I)}_{ij}=\frac{\pi}{N}\exp{\left[\frac{1}{N}A_{ij}A_{ji}\right]}
\]

The product of all those factors together with the normalization
constant $C_{GUE}$ yields the right-hand side of (\ref{gueav}).
\end{proof}
The use of this lemma allows one to represent the ensemble
averaging of the GUE permanental polynomial $p(\mu)$ from
(\ref{start}) in the form
\begin{equation}\label{gue2}
\langle p(\mu)\rangle_{GUE}=\oint \exp{\left[\mu(\xi^*{\bf
z})+\frac{1}{2N}(\xi^*{\bf z})^2\right]}\, \mathcal{D}({\bf
z},\overline{\xi})
\end{equation}
where we used that $Tr({\bf z}\otimes {\xi}^*)^2=(\xi^*{\bf
z})^2$.

Let us now note that the identity (\ref{gueav}) for the particular
case of matrices ${\bf H}$ of the size $1\times 1$ can be written
in the form:
\begin{equation}\label{HS1}
\,\exp{\left[\frac{1}{2N}a^2\right]}=\sqrt{\frac{N}{2\pi}}\int_{-\infty}^{\infty}
e^{-\frac{N}{2}q^2-qa}\, dq
\end{equation}
valid for any complex $a$ and $N>0$ (we used an obvious
replacement $H_{ii}\to q,\,A_{ii}\to a$ and a simple rescaling of
the integration variable with parameter $N$). We are going to use
this formula for $a=(\xi^*{\bf z})$, in order to replace the
factor $\exp{\left[\frac{1}{2N}(\xi^*{\bf z})^2\right]}$ in the
integrand of (\ref{gue2}) with the corresponding Gaussian
integral. Such a procedure (and its generalizations, see below
(\ref{HS2})) is customarily referred to as "the
Hubbard-Stratonovich transformation" in physics\cite{Fyo}.

 Changing the order of
integration, and exploiting (\ref{main}) for the case $ {\bf
F}=(\mu-q){\bf 1}_N$ , i.e.
\[
\oint \exp{\left[(\mu-q)(\xi^*{\bf z})\right]}\, \mathcal{D}({\bf
z},\xi^*)=Per [(\mu-q){\bf 1}_{N}]=(\mu-q)^N
\]
we find
\begin{equation}\label{gue3}
\langle p(\mu)\rangle_{GUE}=
\sqrt{\frac{N}{2\pi}}\int_{-\infty}^{\infty}
e^{-\frac{N}{2}q^2}\,(\mu-q)^N\, dq
\end{equation}
which in view of (\ref{intrep}) is indeed equivalent to
(\ref{GUE0}).

A nice feature of the present method is that it provides, after a
simple generalization, an access to higher-order correlation
functions of permanental polynomials for GUE matrices. In fact, we
will be able to reduce for any integer $n=1,2,\ldots$ the general
$n-$ point correlation function of permanental polynomials of GUE
matrices to the form (\ref{maingau}), and in this way to prove the
{\bf Theorem (\ref{npointgue})}. Note, that
 the volume element $d{\bf q}$ in (\ref{maingau}) is defined in the way analogous to
the measure $d{\bf H}$ in (\ref{GUEmeas}).
\begin{proof}
 We consider in detail the calculation of the two-point
correlation function ($n=2$) defined according to (\ref{main}) as
\begin{eqnarray}\label{gue4}
\langle p(\mu_1)\,p(\mu_2)\rangle_{GUE}=\oint\oint
\exp{\left[\mu_1(\xi_1^*{\bf z_1})+\mu_2(\xi_2^*{\bf
z_2})\right]}\\ \nonumber \times \left\langle\exp{-Tr\left[{\bf
H}\,\left({\bf z_1}\otimes \xi^*_1+{\bf z_2}\otimes
\xi^*_2\right)\right]}\right\rangle_{GUE} \, \mathcal{D}({\bf
z}_1,\xi^*_1)\,\mathcal{D}({\bf z}_2,\xi^*_2)
\end{eqnarray}
The case of general $n$ can be verified precisely along the same
route without any essential modifications.

The ensemble averaging in the right-hand side of (\ref{gue4}) is
performed according to (\ref{gueav}), yielding
\begin{equation}\label{gueav1}
\left\langle\exp{-Tr\left[{\bf H}\,\left({\bf z_1}\otimes
\xi^*_1+{\bf z_2}\otimes
\xi^*_2\right)\right]}\right\rangle_{GUE}= \exp{\left[\frac{1}{2N}
Tr\left({\bf z_1}\otimes \xi^*_1+{\bf z_2}\otimes
\xi^*_2\right)^2\right]}
\end{equation}
Now we notice that
\begin{equation}\label{gueav2}
Tr\left({\bf z_1}\otimes \xi^*_1+{\bf z_2}\otimes
\xi^*_2\right)^2=(\xi_1^*{\bf z_1})^2+(\xi_2^*{\bf
z_2})^2+2(\xi_1^*{\bf z_2})(\xi_2^*{\bf z_1})
\end{equation}
and introducing $2\times 2$ matrix
\begin{equation}\label{gue5}
{\bf a}=\left(\begin{array}{cc}\xi_1^*{\bf z_1} & \xi_1^*{\bf
z_2}\\ \xi_2^*{\bf z_1} & \xi_2^*{\bf z_2}
\end{array}\right)
\end{equation}
we see that
\begin{equation}\label{gue6}
\left\langle\exp{-Tr\left[{\bf H}\,\left({\bf z_1}\otimes
\xi^*_1+{\bf z_2}\otimes
\xi^*_2\right)\right]}\right\rangle_{GUE}=
\exp{\left[\frac{1}{2N}Tr {\bf a}^2\right]}
\end{equation}
Thus
\begin{equation}\label{gue7}
\langle p(\mu_1)\,p(\mu_2)\rangle_{GUE}=\oint\oint
e^{\mu_1(\xi_1^*{\bf z_1})+\mu_2(\xi_2^*{\bf z_2})+\frac{1}{2N}Tr
{\bf a}^2}\, \mathcal{D}({\bf z}_1,\xi^*_1)\,\mathcal{D}({\bf
z}_2,\xi^*_2)
\end{equation}

At the next step we again exploit the Hubbard-Stratonovich
transformation (cf. (\ref{HS1})) :
\begin{equation}\label{HS2}
\exp{\left[\frac{1}{2N}Tr {\bf a}^2\right]}=\int
e^{-\frac{N}{2}Tr{\bf q}^2-Tr({\bf qa})}d{\bf q}
\end{equation}
where ${\bf q}$ is $2\times 2$ Hermitian matrix
\begin{equation}\label{gue8}
{\bf q}=\left(\begin{array}{cc}q_{11} & q^{(R)}_{12}+iq^{(I)}_{12}\\
q^{(R)}_{12}-iq^{(I)}_{12} & q_{22}
\end{array}\right),\quad d{\bf q}=\frac{N^2}{2\pi^2}\,\,dq_{11}\,dq_{22}\,dq^{(R)}_{12}\,dq^{(I)}_{12}
\end{equation}
The relation (\ref{HS2}) follows immediately from (\ref{gueav})
specified for $2\times 2$ matrices, after an appropriate change of
notations. Substituting (\ref{HS2}) back to (\ref{gue7}) and
changing the order of integrations, we arrive at
\begin{equation}\label{gue9}
\langle p(\mu_1)\,p(\mu_2)\rangle_{GUE}=\int e^{-\frac{N}{2}Tr{\bf
q}^2}d{\bf q}\oint\oint e^{\mu_1(\xi_1^*{\bf
z_1})+\mu_2(\xi_2^*{\bf z_2})-Tr ({\bf qa})}\, \mathcal{D}({\bf
z}_1,\xi^*_1)\,\mathcal{D}({\bf z}_2,\xi^*_2)
\end{equation}
Now we note that
\begin{eqnarray}\nonumber
&&\mu_1(\xi_1^*{\bf z_1})+\mu_2(\xi_2^*{\bf z_2})-Tr ({\bf qa})\\
\label{gue10} \,\,& & =(\mu_1-q_{11})(\xi_1^*{\bf
z_1})-q_{12}(\xi_2^*{\bf z_1})-\overline{q}_{11}(\xi_1^*{\bf
z_2})+(\mu_2-q_{22})(\xi_2^*{\bf z_2})\\ \nonumber &\equiv&
({\xi}^*_1,{\xi}^*_2)\left(\begin{array}{cc}(\mu_1-q_{11}){\bf 1}_N & -\overline{q}_{12}{\bf 1}_N \\
-q_{12}{\bf 1}_N  & (\mu_2-q_{22}){\bf 1}_N
\end{array}\right)\left(\begin{array}{c}{\bf z}_1\\ {\bf
z}_2\end{array}\right)
\end{eqnarray}
which allows us to evaluate the contour integrals in (\ref{gue9})
with help of the formula (\ref{main}). The result is simply
\begin{eqnarray}\label{gue11}
&&\oint\oint e^{\mu_1(\xi_1^*{\bf z_1})+\mu_2(\xi_2^*{\bf z_2})-Tr
({\bf qa})}\, \mathcal{D}({\bf z}_1,\xi^*_1)\,\mathcal{D}({\bf
z}_2,\xi^*_2) \\ \nonumber
&=&Per\,\left(\begin{array}{cc}(\mu_1-q_{11}){\bf 1}_N & -\overline{q}_{12}{\bf 1}_N \\
-q_{12}{\bf 1}_N  & (\mu_2-q_{22}){\bf 1}_N
\end{array}\right)
\end{eqnarray}
Using that (i) the permanent of a block-diagonal matrix is equal
to the product of the permanents of the diagonal blocks and (ii)
the matrix transposition does not change its permanent we finally
arrive at the required identity:
\begin{equation}\label{gue99}
\langle p(\mu_1)\,p(\mu_2)\rangle_{GUE}=\int e^{-\frac{N}{2}Tr{\bf
q}^2}\,\left[Per\left({\bf M}-{\bf q}\right)\right]^N\,d{\bf q}
\end{equation}
where ${\bf M}=diag(\mu_1,\mu_2)$.
\end{proof}

According to the Theorem (\ref{npointgue}) the evaluation of the
correlation functions of GUE permanental polynomials reduces to
evaluation of the integral in the right-hand side of
(\ref{maingau}). In the earlier studied case of determinantal
polynomials performing a similar integration was achieved by
shifting the integration variables ${\bf M}-{\bf q}\to {\bf q} $,
and passing to the eigenvalue decomposition ${\bf q}={\bf U}_n{\bf
q}_d{\bf U}_n^*,\, {\bf q}_d=diag(q_1,\ldots,q_n),\, {\bf U}_n\in
U(n)$. As $\det{\bf q}=\det {\bf q_d}$ (and is therefore
independent of ${\bf U}_n$) one can evaluate the unitary-group
integral with the help of HCIZ identity which results in a great
simplification. Unfortunately, permanents lack the nice invariance
properties of the determinants, and for general $n>2$ the exact
evaluation of the corresponding integrals is still outstanding.
The case $n=2$ turns out however to be specific, and a simple
modification of the procedure described above does yield the full
solution of the problem.

To this end, we note that for $2\times 2$ Hermitian matrices {\bf
q} specified in (\ref{gue8})
\[
Per\left({\bf M}-{\bf
q}\right)=(\mu_1-q_{11})(\mu_2-q_{22})+\overline{q}_{12}q_{12}
\]
\[
=(-1)\det{\left(\begin{array}{cc}(-\mu_1+q_{11})& -\overline{q}_{12}\\
-q_{12} & (\mu_2-q_{22})\end{array}\right)}
\]
Substituting this expression to the right-hand side of
(\ref{gue99})), and changing $q_{11}\to -q_{11}$ (which obviously
leaves $Tr{\bf q}^2$ invariant), we therefore see that
\begin{equation}\label{gue111}
\langle p(\mu_1)\,p(\mu_2)\rangle_{GUE}=(-1)^N\int
e^{-\frac{N}{2}Tr{\bf q}^2}\,\left[\det{\left({\bf \tilde{M}}-{\bf
q}\right)}\right]^N\,d{\bf q}
\end{equation}
where ${\bf\tilde{M}}=diag(-\mu_1,\mu_2)$. Now, as explained
above, we can shift the integration variables ${\bf\tilde{
M}}-{\bf q}\to {\bf q} $, and pass to the eigenvalue decomposition
${\bf q}={\bf U}_2{\bf q}_d{\bf U}_2^*,\, {\bf
q}_d=diag(q_1,q_2),\, {\bf U}_2\in U(2)$. The measure is changed
as  $d{\bf q}\to (N^2/4\pi)(q_1-q_2)^2 dq_1\,dq_2\,d\mu_2(U)$, and
\[ Tr{\bf q}^2\to Tr({\bf \tilde{M}}-{\bf
q})^2=(\mu_1^2+\mu_2^2)-2Tr\left({\bf \tilde{M}}{\bf U}_2{\bf
q}_d{\bf U}_2^* \right)+(q_1^2+q_2^2).\] Combining all these
facts, we find that
\begin{eqnarray}\label{gue12}
&&\langle p(\mu_1)\,p(\mu_2)\rangle_{GUE}=\frac{N^2}{4\pi}
e^{-\frac{N}{2} ( \mu_1^2+\mu_2^2)} \\ \nonumber &&\times
\int_{-\infty}^{\infty} \int_{-\infty}^{\infty} e^{-\frac{N}{2} (
q_1^2+q_2^2)}(q_1q_2)^N (q_1-q_2)^2 dq_1dq_2 \, I_{HCIZ}(q_1,q_2)
\end{eqnarray}
where we defined
\begin{eqnarray}\nonumber
I_{HCIZ}(q_1,q_2)&=&\int \exp{-NTr\left({\bf \tilde{M}}{\bf
U}_2{\bf q}_d{\bf U}_2^* \right)}\,d\mu_2(U)\\ \label{IZHC2}
&=& \frac{1}{N(q_1-q_2)(\mu_1+\mu_2)}\det{\left(\begin{array}{cc} e^{N\mu_1q_1}& e^{-N\mu_2q_1}\\
e^{N\mu_1q_2} & e^{-N\mu_2q_2}\end{array}\right)}
\end{eqnarray}
according to (\ref{IZHCfull}) as the standard HCIZ integral. After
straightforward algebraic manipulations the formulae
(\ref{gue12})-(\ref{IZHC2})) give rise to the following final
expression for the two-point correlation function:
\begin{equation}\label{finalgue}
\langle p(\mu_1)\,p(\mu_2)\rangle_{GUE}=
\frac{1}{(-i\mu_1)-(i\mu_2)}
\det{\left(\begin{array}{cc} \pi_N(-i\mu_1)& \pi_N(i\mu_2)\\
\pi_{N+1}(-i\mu_1) & \pi_{N+1}(i\mu_2)\end{array}\right)},
\end{equation}
where we have used the integral representation (\ref{intrep}) for
the monic Hermite polynomials.

In particular, last expression proves a curious relation
(\ref{relgue}) between correlation functions of permanental and
characteristic polynomials of GUE matrices,  valid in view of the
corresponding Brezin-Hikami result (\ref{bh2}).

\section{Two-point correlation function of the permanental polynomials
for matrices from the Circular Unitary Ensemble} \label{Section5}

Consider the space $\mathcal{U}$ (identified with the unitary
group $U(N)$) of $N\times N$ unitary matrices,
 satisfying ${\bf
U}^{-1}={\bf U}^*$, with the probability measure on $\mathcal{U}$
being chosen according to the Haar's measure on the unitary group.
This condition defines the Circular Unitary Ensemble (CUE). With
the permanental polynomial defined as $p(\mu)=Per (\mu{\bf
1}_N-{\bf U})$, it is immediately clear that any non-trivial
correlation function must have the form of a product of an even
number of permanental polynomials, precisely half of the factors
involving matrices ${\bf U}^*$, that is
\begin{equation}\label{corCUE}
\left\langle
\prod_{k=1}^n\,p\left(\mu^{(A)}_k\right)\overline{p\left(\mu^{(B)}_k\right)}\right\rangle_{CUE}
\end{equation}
In what follows we will address the simplest non-trivial case of
the two-point function ( $n=1$), which according to
(\ref{contint}) has the following integral representation:
\begin{eqnarray}\label{cue1}
\nonumber\left\langle
p(\mu^{(A)})\,\overline{p(\mu^{(B)})}\right\rangle_{CUE}&=&\oint\oint
e^{\left[\mu^{(A)}(\xi_A^*{\bf z_A})+\overline{\mu^{(B)}}(\xi_B^*{\bf z_B})\right]}\\
 &\times& \left\langle e^{-Tr\left[{\bf U}\,\left({\bf
z_A}\otimes \xi^*_A)+{\bf U}^*({\bf z_B}\otimes
\xi^*_B\right)\right]}\right\rangle_{CUE} \, \mathcal{D}({\bf
z}_A,\xi^*_A)\,\mathcal{D}({\bf z}_B,\xi^*_B)
\end{eqnarray}
To perform the ensemble average of the exponential factor in the
integrand above we make use of the following
\begin{thm}\label{FKthm}
For two general complex square $N\times N$ matrices ${\bf A}$ and
${\bf B}$ define
\begin{equation}\label{Fdef}
F({\bf A}{\bf B}^*)=\int_{U(N)} \exp{Tr\left[{\bf A}{\bf U}+{\bf
U}^*{\bf B}^*\right]}\,\,d\mu_N(U)
\end{equation}
Suppose ${\bf AB}^*$ has rank 1. Then for $N\ge 2$ holds
\begin{equation}\label{Fdef1}
F({\bf A}{\bf B}^*)=(N-1)\int_0^1 (1-t)^{N-2}
I_0(2\sqrt{tv^2})\,\,dt
\end{equation}
where $v^2$ is the only non-zero eigenvalue of ${\bf AB}^*$ and
\begin{equation}\label{Bessel}
I_0(z)=\sum_{j=0}^{\infty}\frac{z^{2j}}{2^{2j}j!^2}
\end{equation}
is the modified Bessel function.
\end{thm}
\begin{proof}
The theorem is just a special rank$-1$ case of a more general
statement proved for any rank $2m\le N$ in \cite{FK} using the
Schur function expansion. It is intimately related to the
so-called "bosonic colour-flavour" transformation introduced by
Zirnbauer in \cite{zirn}. For rank $m=N$ the function $F({\bf
A}{\bf B}^*)$ was earlier calculated in \cite{SW}, see also
\cite{BA}.
\end{proof}
According to (\ref{cue1}) we are interested in a particular case
${\bf A}=-{\bf z_A}\otimes \xi_A^* $ and ${\bf B}^*=-{\bf
z_B}\otimes \xi_B^*$, so that ${\bf A B}^*=(\xi_A^*{\bf
z_B})\,{\bf z_A}\otimes \xi_B^* $, which is obviously of rank 1,
with $v^2=(\xi_A^*{\bf z_B})(\xi_B^*{\bf z_A})$. Using
(\ref{Fdef1}) and the expansion (\ref{Bessel}) it is convenient to
perform integration over $t$ explicitly, and find:
\begin{equation}
\left\langle e^{-Tr\left[{\bf U}\,\left({\bf z_A}\otimes
\xi^*_A)+{\bf U}^*({\bf z_B}\otimes
\xi^*_B\right)\right]}\right\rangle_{CUE}
=(N-1)!\sum_{j=0}^{\infty}\frac{\left[(\xi_A^*{\bf
z_B})(\xi_B^*{\bf z_A})\right]^{j}}{j!(N-1+j)!}
\end{equation}
Substituting this expression to (\ref{cue1}) we need to evaluate
the integral
\begin{eqnarray}\label{cue4}
&&I_j(\mu^{(A)},\overline{\mu^{(B)}})\\ &=&\nonumber\oint\oint
e^{\left[\mu^{(A)}(\xi_A^*{\bf
z_A})+\overline{\mu^{(B)}}(\xi_B^*{\bf
z_B})\right]}\left[(\xi_A^*{\bf z_B})(\xi_B^*{\bf z_A})\right]^{j}
\mathcal{D}({\bf\ z}_A,\xi^*_A)\,\mathcal{D}({\bf z}_B,\xi^*_B)
\end{eqnarray}
This is most easily done by first considering the generating
function
\begin{eqnarray}\label{cue5}
&&I_j(\mu^{(A)},\overline{\mu^{(B)}},p_1,p_2)\\
&=&\nonumber\oint\oint e^{\left[\mu^{(A)}(\xi_A^*{\bf
z_A})+\overline{\mu^{(B)}}(\xi_B^*{\bf z_B})+p_1(\xi_A^*{\bf
z_B})+p_2(\xi_B^*{\bf z_A})\right]} \mathcal{D}({\bf\
z}_A,\xi^*_A)\,\mathcal{D}({\bf z}_B,\xi^*_B)
\end{eqnarray}
from which (\ref{cue4}) is obtained by a simple differentiation,
and then taking limit $p_1=p_2=0$. Noticing
\begin{eqnarray}
&&\mu^{(A)}(\xi_A^*{\bf z_A})+\overline{\mu^{(B)}}(\xi_B^*{\bf
z_B})+p_1(\xi_A^*{\bf z_B})+p_2(\xi_B^*{\bf z_A})\\
\nonumber &\equiv&
({\xi}^*_A,{\xi}^*_B)\left(\begin{array}{cc}\mu^{(A)}{\bf 1}_N & p_1{\bf 1}_N \\
p_2{\bf 1}_N  & \overline{\mu^{(B)}}{\bf 1}_N
\end{array}\right)\left(\begin{array}{c}{\bf z}_A\\ {\bf
z}_B\end{array}\right)
\end{eqnarray}
we can apply the Lemma (\ref{contint}) to (\ref{cue5}) and find
\begin{equation}\label{cue6}
I_j(\mu^{(A)},\overline{\mu^{(B)}},p_1,p_2)=\left[Per\left(\begin{array}{cc}\mu^{(A)}& p_1\\
p_2  & \overline{\mu^{(B)}}
\end{array}\right)\right]^N,
\end{equation}
where we have used that the permanent of a block-diagonal matrix
is equal to the product of the permanents of the diagonal blocks.
Now a simple differentiation yields the required result:
\begin{eqnarray}\label{cue7}
I_j(\mu^{(A)},\overline{\mu^{(B)}})=\left\{\begin{array}{c}\frac{N!j!}{(N-j)!}(\mu^{(A)}\overline{\mu^{(B)}})^{N-j},\quad
j\le N\\ 0,\quad\quad\quad\quad\quad\quad\quad\quad\quad
j>N\end{array}\right.
\end{eqnarray}
Combining all these results, and changing $(N-j)\to j$ for
convenience we arrive at the final expression for the two-point
correlation function (\ref{cue1}):
\begin{equation}\label{cue8}
\left\langle
p(\mu^{(A)})\,\overline{p(\mu^{(B)})}\right\rangle_{CUE}= N!(N-1)!
\sum_{j=0}^N
\frac{(\mu^{(A)}\overline{\mu^{(B)}})^{j}}{j!(2N-1-j)!}
\end{equation}
A direct check shows that the above expression can also be written
in the form of an integral:
\begin{equation}\label{saddle1}
\left\langle
p(\mu^{(A)})\,\overline{p(\mu^{(B)})}\right\rangle_{CUE}=
(N-1)\int_0^1 (1-t)^{N-2}(\mu^{(A)}\overline{\mu^{(B)}}+t)^N\,dt
\end{equation}

This latter form is especially suitable for extracting the
asymptotic behaviour in the limit of large $N$ by the Laplace
method. In particular, for $\mu^{(A)}=\mu^{(B)}=\mu$ the integral
is dominated by the vicinity of the point of maximum of the
integrand $t=t_m=(1-|\mu|^2)/2$ for $|\mu|<1$, and by the lower
limit of integration $t=0$ for $|\mu|\ge 1$. A straightforward
calculation then shows that
\begin{equation}\label{saddle2}
\Phi_p(|\mu|^2)=\lim_{N\to\infty}\frac{1}{N}\ln{\left\langle
|p(\mu)|^2\right\rangle_{CUE}}=\left\{\begin{array}{c}
2\ln{\left(\frac{1+|\mu|^2}{2}\right)}, \quad |\mu|<1\\
 \ln{|\mu^2|}, \quad |\mu| \ge 1\end{array} \right.
\end{equation}
 Under assumption that the relation (\ref{commute}) holds for CUE
matrices, the expression (\ref{saddle2}) for $\Phi_p$ yields after
applying to it the Laplace operator a two-dimensional density
profile in the complex plane $\mu=x+iy$ given in the equation
(\ref{dencomp1}).

Since this result is very different from the density of
(unimodular) eigenvalues of the CUE matrices, let us briefly show
how to recover the latter density in a similar approach. The
analogue of (\ref{saddle1}) was found in \cite{FK}, and reads
\begin{equation}\label{saddle3}
\left\langle
d(\mu^{(A)})\,\overline{d(\mu^{(B)})}\right\rangle_{CUE}=
(N+1)\int_0^\infty
\frac{1}{(1+t)^{N+2}}(\mu^{(A)}\overline{\mu^{(B)}}+t)^N \,dt
\end{equation}
where $d(\mu)=\det{(\mu {\bf 1}_N-{\bf U})}$ is the corresponding
characteristic polynomial. In contrast to the previously
considered case, in the limit of large $N$ the integrand for
$\mu^{(A)}=\mu^{(B)}=\mu$ does not have a maximum in the domain of
integration, and thus the integral is dominated by the vicinity of
the upper limit $t=\infty$ for $|\mu|<1$, and by the vicinity of
the lower limit $t=0$ for $|\mu|>1$. This yields
\begin{equation}\label{saddle4}
\Phi_d(|\mu|^2)=\lim_{N\to\infty}\frac{1}{N}\ln{\left\langle
|d(\mu)|^2\right\rangle_{CUE}}=\left\{\begin{array}{c}
0,\quad \quad |\mu|<1\\
 \ln{|\mu^2|}, \quad |\mu|>1\end{array} \right.
\end{equation}
We see, that the function $\Phi_d(x,y)$ is harmonic everywhere in
the complex plane $\mu=x+iy$, except the unit circle $|\mu|=1$.
This means that the density of eigenvalues of CUE matrices is
indeed vanishing everywhere except $|z|=1$, in agreement with
obvious unimodularity of the corresponding eigenvalues. The
discontinuity of the normal derivative of the "potential" $\Phi_1$
across the unit circle yields the constant density on the circle,
as expected.

\section{Correlation functions for permanental polynomials of
Ginibre  matrices}\label{section6}

The joint probability density of matrix entries for complex
Ginibre $N\times N$ matrices ${\bf Z}$ is given by
\begin{equation}\label{Ginmeas}
\mathcal{P}_{GUE}({\bf Z})d{\bf Z}\,d{\bf Z}^*=C_{Gin}
\prod_{i,j}^N e^{-N \overline{Z}_{ij}Z_{ij}}
\prod_{i,j}^N\,dZ^{R}_{ij}dZ^{I}_{ij}
\end{equation}
where $C_{Gin}=\left(N/\pi\right)^{N^2}$ is the corresponding
normalization constant.
\begin{lem}
 The identity
\begin{equation}\label{ginav}
\left\langle e^{-Tr ({\bf Z A}+{\bf B Z^*})}
\right\rangle_{Gin}=e^{\frac{1}{N}Tr{\bf AB}}
\end{equation}
holds for any two general $N\times N$ matrices ${\bf A,\,B}$ with
complex entries.
\end{lem}
\begin{proof}
The integral in the left hand side factorizes into a product of
simple Gaussian integrals
\[
\int
e^{-N\overline{Z}_{ij}Z_{ij}-Z_{ij}A_{ji}-\overline{Z}_{ij}B_{ij}}
dZ^{(R)}_{ij}dZ^{(I)}_{ij}
=\frac{\pi}{N}\,e^{\frac{1}{N}A_{ji}B_{ij}}
\]
The product of all those factors together with the normalization
constant $C_{GUE}$ yields the right-hand side of (\ref{ginav}).\\

\end{proof}

Again, as in the case of CUE the nontrivial correlation functions
for Ginibre ensemble must contain even number of factors, with
half of them being permanental polynomials of $Z$, the rest being
permanental polynomials of $Z^*$. More precisely, we will prove
the following
\begin{thm}
For any integer $n=1,2,\ldots$ the $2n-$ point correlation
function of permanental polynomials of Ginibre matrices ${\bf Z}$
has the following integral representation:
\begin{equation}\label{maingin}
\left\langle
\prod_{k=1}^n\,p\left(\mu^{(A)}_k\right)\overline{p\left(\mu^{(B)}_k\right)}\right\rangle_{Gin}=\int
e^{-N\,Tr{\bf Q}^2}\,\left[Per\left(\begin{array}{cc}{\bf M}_A&
{\bf Q}^*\\{\bf Q}&{\bf M}_B\end{array}\right)\right]^N\,d{\bf
Q}\,d{\bf Q}^*
\end{equation}
In this expression ${\bf
M}_{A}=diag(\mu^{(A)}_1,\ldots,\mu^{(A)}_n)$ and ${\bf
M}_{B}=diag(\overline{\mu}^{(B)}_1,\ldots,\overline{\mu}^{(B)}_n)$
are $n\times n$ diagonal matrices and ${\bf Q}$ is a general
$n\times n$ complex matrix, with the volume element $d{\bf
Q}\,d{\bf Q}^* $ defined in the way analogous to the measure
$d{\bf Z}\,d{\bf Z}^* $ in (\ref{Ginmeas}).
\end{thm}
\begin{proof}
 We consider in detail the calculation of the two-point
correlation function ($n=1$) defined according to (\ref{main}) as
\begin{eqnarray}\label{gin4}
&&\left\langle
p\left(\mu^{(A)}\right)\overline{p\left(\mu^{(B)}\right)}\right\rangle_{Gin}=\oint\oint
\exp{\left[\mu^{(A)}(\xi_A^*{\bf
z_A})+\overline{\mu}^{(B)}(\xi_B^*{\bf z_B})\right]}\\ \nonumber
&&\times \left\langle\exp{-Tr\left[{\bf Z}\,{\bf z_A}\otimes
\xi^*_A+{\bf Z}^*{\bf z_B}\otimes
\xi^*_B\right]}\right\rangle_{Gin} \, \mathcal{D}({\bf
z}_A,\xi^*_A)\,\mathcal{D}({\bf z}_B,\xi^*_B)
\end{eqnarray}

The ensemble averaging in the right-hand side of (\ref{gue4}) is
performed according to (\ref{ginav}), and using $Tr\left({\bf
z_A}\otimes \xi^*_A\right)\left({\bf z_B}\otimes
\xi^*_B\right)=\left(\xi^*_A{\bf z_B}\right)\left(\xi^*_B{\bf
z_A}\right)$
 yielding
\begin{equation}\label{gin5}
\left\langle\exp{-Tr\left[{\bf Z}\,{\bf z_A}\otimes \overline
{\xi}_A+{\bf Z}^*{\bf z_B}\otimes \overline
{\xi}_B\right]}\right\rangle_{Gin}= \exp{\left[\frac{1}{N}
\left(\xi^*_A{\bf z_B}\right)\left(\xi^*_B{\bf z_A}\right)\right]}
\end{equation}
Introducing the notations
\begin{equation}\label{PP}
P_{AB}=\left(\xi^*_A{\bf z_B}\right),\, P_{BA}=\left(\xi^*_B{\bf
z_A}\right)
\end{equation}
we notice that the relevant form of the Hubbard-Stratonovich
transformation in this case is given by (cf. (\ref{ginav}))
\begin{equation}\label{HSgin1}
e^{\frac{1}{N}\left[P_{AB}P_{BA}\right]}=\frac{N}{\pi}\int\int
e^{-N\overline{Q}Q+QP_{AB}+\overline{Q}P_{BA}} dQ^{(R)}dQ^{(I)}
\end{equation}
where the integral over $Q$ is in the plane of a general complex
variable, $Q=Q^{R}+iQ^{I}$. Substituting this back to (\ref{gin5})
and then to (\ref{gin4}) and changing the order of integrations,
we have
\begin{eqnarray}
\nonumber &&\left\langle
p\left(\mu^{(A)}\right)\overline{p\left(\mu^{(B)}\right)}\right\rangle_{Gin}=
\frac{N}{\pi}\int\int e^{-N\overline{Q}Q} \,dQ\,d\overline{Q}\\
\nonumber &&\times\oint\oint \exp{\left[\mu^{(A)}(\xi_A^*{\bf
z_A})+\overline{\mu}^{(B)}(\xi_B^*{\bf z_B})+Q\,(\xi^*_A{\bf z_B})
+\overline{Q}(\xi^*_B{\bf z_A})\right]} \, \mathcal{D}({\bf
z}_B,\xi^*_B)\,\mathcal{D}({\bf z}_B,\xi^*_B).
\end{eqnarray}
Further noticing that
\[
\mu^{(A)}(\xi_A^*{\bf z_A})+\overline{\mu}^{(B)}(\xi_B^*{\bf
z_B})+Q\,(\xi^*_A{\bf z_B}) +\overline{Q}(\xi^*_B{\bf
z_A})=(\xi^*_A,\xi^*_B)\left(\begin{array}{cc} \mu^{(A)}{\bf 1}_N&
\overline{Q}{\bf 1}_N\\Q{\bf 1}_N &\overline{\mu}^{(B)}{\bf
1}_N\end{array}\right)\left(\begin{array}{c}{\bf z}_A\\ {\bf z}_B
\end{array}\right)
\]
the contour integral can be performed using (\ref{main}), yielding
finally
\begin{eqnarray}
  \nonumber\left\langle
p\left(\mu^{(A)}\right)\overline{p\left(\mu^{(B)}\right)}\right\rangle_{Gin}&=&
\frac{N}{\pi}\int\int
e^{-N\overline{Q}Q}\,Per\left(\begin{array}{cc} \mu^{(A)}{\bf
1}_N& \overline{Q}{\bf 1}_N\\Q{\bf 1}_N
&\overline{\mu}^{(B)}{\bf 1}_N\end{array}\right)\,dQ \,d\overline{Q}\\
  &=& \frac{N}{\pi}\int\int
e^{-N\overline{Q}Q}\,\left[Per\left(\begin{array}{cc} \mu^{(A)}&
\overline{Q}\\Q&\overline{\mu}^{(B)}\end{array}\right)\right]^N\,
dQ\,d\overline{Q} \label{fingin}
\end{eqnarray}
in accordance with (\ref{maingin}).\\

The proof for general $n$ follows precisely the same lines with
minimal changes. The place of two parameters $P_{AB}$ and $P_{BA}$
defined in (\ref{PP}) is taken by two $n\times n$ matrices ${\bf
P}_{AB}$ and ${\bf P}_{BA}$ with entries
\begin{equation}\label{AB1}
\left[P_{AB}\right]_{ij}=\left(\xi_{A,i}^*{\bf z}_{B,j}\right),\,
\left[P_{BA}\right]_{ij}=\left(\xi^*_{B,i}{\bf z}_{A,j}\right)
\end{equation}
where $i,j=1,\ldots,n$, and the corresponding Hubbard-Stratonovich
transformation is
\begin{equation}\label{HSgin2}
e^{\frac{1}{N}Tr\left[{\bf P}_{AB}{\bf P}_{BA}\right]} =\int
e^{-N\,Tr\left[{\bf Q}^*{\bf Q}\right]+Tr\left[{\bf QP}_{AB}+{\bf
Q}^*{\bf P}_{BA}\right]} d{\bf Q}\,d{\bf Q}^*
\end{equation}
\end{proof}

{\bf Note}: The problem of evaluating the integral in the
right-hand side of (\ref{maingin}) for general $n>1$ is still
outstanding. For $n=1$ the calculation is elementary:
\begin{eqnarray}\nonumber
\left\langle
p\left(\mu^{(A)}\right)\overline{p\left(\mu^{(B)}\right)}\right\rangle_{Gin}
  &=& \frac{N}{\pi}\int\int
e^{-N\overline{Q}Q}\,\left[\mu^{(A)}\overline{{\mu}}^{(B)}+\overline{Q}Q\right]^N\,
dQ\,d\overline{Q} \label{fingin1}\\ \label{elemgin}=&&
N\int_0^{\infty}
e^{-NR}\left(\mu^{(A)}\overline{{\mu}}^{(B)}+R\right)^N\,dR\\
=&&\frac{N!}{N^N}\sum_{k=0}^N\frac{1}{k!}\left[N\mu^{(A)}\overline{{\mu}}^{(B)}\right]^k
\end{eqnarray}
The integral form (\ref{elemgin}) is most convenient for
extracting the large-N asymptotics by the Laplace method. For
$\mu^{(A)}=\mu^{(B)}=\mu$ the integrand has a sharp maximum around
$R=1-|\mu|^2$ as long as $|\mu|<1$ and is dominated by the lower
limit $R=0$ for $|\mu|>1$. This gives:
\begin{equation}\label{saddle5}
\Phi_{Gin}(|\mu|^2)=\lim_{N\to\infty}\frac{1}{N}\ln{\left\langle
|p(\mu)|^2\right\rangle_{Gin}}=\left\{\begin{array}{c}
|\mu|^2-1, \quad |\mu|<1\\
 \ln{|\mu^2|}, \quad |\mu| \ge 1\end{array} \right.
\end{equation}
Under further assumption of validity of (\ref{commute}) such an
expression implies the uniform density of permanental roots inside
the unit circle, as conjectured in (\ref{dencomp2}).

\section{Correlation functions for permanental polynomials of GOE
matrices}\label{section7}

The joint probability density of matrix entries for GOE $N\times
N$ real symmetric matrices ${\bf H}$ is given by
\begin{equation}\label{GOEmeas}
\mathcal{P}_{GOE}({\bf H})d{\bf H}=C_{GOE} \prod_{i=1}^N
e^{-\frac{N}{2}H_{ii}^2}\prod_{i<j} e^{-N H^2_{ij}}
\prod_{i=1}^NdH_{ii}\prod_{i<j}\,dH_{ij}
\end{equation}
where $C_{GOE}=2^{-\frac{N}{2}}\left(N/\pi\right)^{N(N+1)/4}$ is
the corresponding normalization constant.

It is straightforward to verify the following
\begin{lem}
 The identity
\begin{equation}\label{goeav}
\left\langle e^{-Tr {\bf H A}}
\right\rangle_{GOE}=e^{\frac{1}{4N}\left[Tr{\bf A}^2+Tr{\bf
AA}^T\right]}
\end{equation}
holds for any general $N\times N$ matrix ${\bf A}$ with complex
entries.
\end{lem}

The use of this lemma allows one to represent the ensemble
averaging of the GOE permanental polynomial $p(\mu)$ from
(\ref{start}) in the form
\begin{equation}\label{goe2}
\langle p(\mu)\rangle_{GOE}=\oint \exp{\left[\mu(\xi^*{\bf
z})+\frac{1}{4N}\left(\sum_i^N\overline{\xi}_iz_i\right)^2+\frac{1}{4N}
\sum_i^N\overline{\xi}^2_i\sum_i^Nz_i^2\right]}\, \mathcal{D}({\bf
z},\overline{\xi})
\end{equation}
where we used that $Tr({\bf z}\otimes {\xi}^*)({\bf z}\otimes
{\xi}^*)^T=\sum_i\overline{\xi}^2_i\sum_iz_i^2$.

Following the same strategy as for GUE case, we use the
appropriate gaussian integrals ("Hubbard-Stratonovich
transformation") to linearize the quartic terms in the
exponential. The first term is identical (up to the factor $1/2$)
to that in the GUE case, and we can use (\ref{HS1}). The second
term in the exponential is dealt with exploiting (\ref{HSgin1}),
with the correspondence
$P_{AB}\to\sum_i\overline{\xi}^2_i,\,P_{BA}\to\sum_iz_i^2$.
Exchanging the order of integrations, and remembering that
$\mathcal{D}({\bf z},\xi^*)$ is essentially given by
$\frac{1}{(2\pi)^{2N}}\prod_{k=1}^N\frac{dz_k}{z_k^2}\prod_{k=1}^N
\frac{d\overline{\xi}_k}{\overline{\xi}_k^2}$ (see
(\ref{shorthand})), we arrive at
\begin{eqnarray}\label{goegoe}
\langle
p(\mu)\rangle_{GOE}=\sqrt{\frac{N}{2\pi}}\int_{-\infty}^{\infty}
e^{-\frac{N}{2}q^2}\, dq\, \frac{N}{\pi}\int e^{-N\overline{q_1}q_1}d{\overline q_1}dq_1\\
\nonumber \times \left[
\frac{1}{(2\pi)^2}\oint_{|z|=1}\oint_{|\xi|=1}
e^{\left[(\mu-\frac{1}{\sqrt{2}}q)\overline{\xi}z-\frac{1}{2}q_1z^2-\frac{1}{2}\overline{q_1}\overline{\xi}^2\right]}\,\,
\frac{dz}{z^2}\frac{d\overline{\xi}}{\overline{\xi}^2}\right]^N
\end{eqnarray}
Noticing, that the result of the contour integration is
independent of $q_1,\overline{q}_1$ (as any analytic factors
depending on $z^2$ or $\overline{\xi}^2$ could be replaced by
zero's term in their Taylor expansion) we see that second line in
(\ref{goegoe}) obviously produces $(\mu-\frac{1}{\sqrt{2}}q)^N$,
and after simple manipulations and exploitation of (\ref{intrep})
we arrive at the following
\begin{prop}
The expectation value of the permanental polynomial for the
Gaussian Orthogonal Ensemble (GOE) is given in terms of the monic
Hermite polynomial (\ref{intrep}) by
\begin{equation}\label{GOE0}
\langle
p\,(\mu)\rangle_{GOE}=\frac{i^N}{2^{N/2}}\,\pi_{N}(-i\sqrt{2}\mu)=i^N\,\langle
d\,(-i\mu)\rangle_{GOE}\,,
\end{equation}
\end{prop}
where again $d(\mu)=\det{(\mu{\bf 1}_N-H)}$. The second equality
follows from comparison with the equation (48) of \cite{BH3}.

The same method can be used, {\it mutatis mutandis}, for
evaluating higher correlation function, although actual
calculations are much more cumbersome than for the GUE case. Below
we restrict ourselves to considering explicitly the two-point
correlation function, $n=2$. As the starting expression we use the
analogue of (\ref{gue4}):
\begin{eqnarray}\label{goe44}
\langle p(\mu_1)\,p(\mu_2)\rangle_{GOE}=\oint\oint
\exp{\left[\mu_1(\xi_1^*{\bf z_1})+\mu_2(\xi_2^*{\bf
z_2})\right]}\\ \nonumber \times \left\langle\exp{-Tr\left[{\bf
H}\,\left({\bf z_1}\otimes \xi^*_1+{\bf z_2}\otimes
\xi^*_2\right)\right]}\right\rangle_{GOE} \, \mathcal{D}({\bf
z}_1,\xi^*_1)\,\mathcal{D}({\bf z}_2,\xi^*_2)
\end{eqnarray}
where the ensemble average is now performed using the identity
(\ref{goeav}), with the role of ${\bf A}$ played by the matrix
${\bf z_1}\otimes \xi^*_1+{\bf z_2}\otimes \xi^*_2$.

According to (\ref{gueav2}), for such a matrix  $Tr{\bf
A}^2=Tr{\bf a}^2$, where ${\bf a}$ was introduced in (\ref{gue5}).
We also notice that
\[
Tr{\bf A A}^T=\sum_{i=1}^N\overline{\xi}_{1i}^2 \sum_{i=1}^N
z_{1i}^2+\sum_{i=1}^N\overline{\xi}_{2i}^2 \sum_{i=1}^N
z_{2i}^2+2\sum_{i=1}^N\overline{\xi}_{1i}\overline{\xi}_{2i}\sum_{i=1}^N
z_{1i}z_{2i}
\]
Combining Hubbard-Stratonovich transformations (\ref{HS2}) with
(three times) (\ref{HS1}) we see that the ensemble average
featuring in (\ref{goe44}) can be represented as
\begin{eqnarray}\label{goeav1}
\nonumber &&  \left\langle e^{-Tr\left[{\bf H}\,\left({\bf
z_1}\otimes \xi^*_1+{\bf z_2}\otimes
\xi^*_2\right)\right]}\right\rangle_{GOE}=\frac{2N^3}{\pi^3}\int
e^{-NTr{\bf q}^2-Tr({\bf qa})}d{\bf q} \int
e^{-N(\overline{q_1}q_1+\overline{q_2}q_2+2\overline{q_3}q_3)}\\
\nonumber &&\times
e^{\frac{1}{2}\sum_{k=1}^2\left[q_k\sum_{i=1}^Nz_{ki}^2+\overline{q_k}\sum_{i=1}^N\overline{\xi}_{ki}^2
+q_3\sum_{i=1}^Nz_{1i}z_{2i}+\overline{q_3}\sum_{i=1}^N\overline{\xi}_{1i}\overline{\xi}_{1i}\right]}
d{\overline q_1}dq_1d{\overline q_2}dq_2d{\overline q_3}dq_3
\end{eqnarray}
Substituting this back to (\ref{goe44}), changing the order of
integration, and suppressing the terms which do not contribute to
the value of the contour integral, we obtain after a simple
manipulation:
\begin{eqnarray}\label{goe55}
&&\langle p(\mu_1)\,p(\mu_2)\rangle_{GOE}\\
\nonumber && =\frac{2N}{\pi}\int e^{-N\left(Tr{\bf
q}^2+2\overline{q}_3q_3\right)}\left[\mathcal{J}({\bf
q},\overline{q}_3,q_3;\mu_1,\mu_2)\right]^Nd{\bf q} d{\overline
q_3}dq_3
\end{eqnarray}
where $Tr{\bf q}^2=q_{11}^2+q_{22}^2+2\overline{q}_{12}q_{12}$ and
\begin{eqnarray}\label{goe66}
&&\mathcal{J}_N({\bf
q},\overline{q}_3,q_3;\mu_1,\mu_2)=\frac{1}{(2\pi)^{4}}\oint_{|z_1|=1}\oint_{|\xi_1|=1}\oint_{|z_2|=1}\oint_{|\xi_2|=1}
e^{\overline{q}_3\overline{\xi}_1\overline{\xi}_2+q_3z_{1}z_{2}}\\
\nonumber &&\times
\exp{\left[(\overline{\xi}_1,\overline{\xi}_2)\left(\begin{array}{cc}\mu_1-q_{11}&-\overline{q}_{12}\\
-q_{12}&\mu_2-q_{22}\end{array}\right)\left(\begin{array}{c}z_1\\z_2\end{array}\right)\right]
}\prod_{k=1}^2\frac{dz_k}{z_k^2}\prod_{k=1}^2
\frac{d\overline{\xi}_k}{\overline{\xi}_k^2}\\ \nonumber
&&=\left[(\mu_1-q_{11})(\mu_2-q_{22})+\overline{q}_{12}q_{12}+\overline{q}_{3}q_{3}\right]
\end{eqnarray}
 Changing finally $q_{11}\to -q_{11}$ we
arrive at the final expression
\begin{eqnarray}\label{goe77}
\langle p(\mu_1)\,p(\mu_2)\rangle_{GOE} =\frac{2N}{\pi}\int
e^{-N\left(Tr{\bf q}^2+2\overline{q}_3q_3\right)} &&\\ \nonumber
\times
\left[(\mu_1+q_{11})(\mu_2-q_{22})+\overline{q}_{12}q_{12}+\overline{q}_{3}q_{3}\right]^N
d{\bf q} d{\overline q_3}dq_3
\end{eqnarray}
Comparing this result with the equation (43) of \cite{BH3} we
infer the validity of the relation:
\begin{equation}\label{relgoe}
\langle p(\mu_1)\,p(\mu_2)\rangle_{GOE}=\langle
d(i\mu_1)\,d(-i\mu_2)\rangle_{GOE}\,,\quad
\end{equation}
which is precisely the same as for GUE case, (\ref{relgue}).\\

\appendix

\section{Permanents as multivariate Gaussian
integrals}

 We start with verifying the following
\begin{lem}\label{lemapp}
Let ${\bf F}>0$ be positive definite $N\times N$  Hermitian
matrix, ${\bf a}=(a_1,\ldots,a_N)^T, {\bf b}=(b_1,\ldots,b_N)^T,
{\bf s}=(s_1,\ldots,s_N)^T $all be $N-$ component complex vectors,
and $d{\bf s}^*d{\bf s}=\prod_{i=1}^{N} d{\mbox Re}{(s_i)}d{\mbox
Im}{(s_i)}$. Then the following identity holds
\begin{equation}\label{gauapp}
e^{{\bf a}^*{\bf F^{-1}}{\bf b}}=\det{\bf F}\,\int  e^{-{\bf s}^*
{\bf F}{\bf s}-({\bf a}^*{\bf s})-({\bf s}^*{\bf b})} d{\bf
s}^*d{\bf s}
\end{equation}
\end{lem}
\begin{proof}
Representing ${\bf F}={\bf U}{\bf f}{\bf U}^*$ in terms of the
diagonal matrix of the eigenvalues ${\bf
f}=\mbox{diag}(f_1,\ldots,f_N)>0$ and the unitary matrix ${\bf U}$
of the eigenvectors, and introducing new variables of integration
${\bf q}={\bf U}^*{\bf s}$ so that $d{\bf s}^*d{\bf s}=d{\bf
q}^*d{\bf q}$, the  $N-$fold gaussian integral in the right-hand
side  decomposes into the product of $N$ simple Gaussian integrals
of the type (\ref{HSgin1}) :
\begin{equation}
\prod_{i=1}^N \int
e^{-f_i\overline{q}_i{q_i}-\overline{\tilde{a}}_iq_i-\overline{
q}_i\tilde{b}_i} d\overline{q}_id{q_i}=\prod_{i=1}^N
e^{\frac{1}{f_i}\,\overline{\tilde{a}}_i\,\tilde{b}_i}
\end{equation}
where we introduced the notations $\tilde{\bf a}={\bf U}^*{\bf a}$
and $\tilde{\bf b}={\bf U}^*{\bf b}$. We further find that
\begin{equation}
\sum_{i=1}^N
\frac{1}{f_i}\,\overline{\tilde{a}}_i\,\tilde{b}_i=\sum_{jk}\overline{a}_j
\left(\sum_{i=1}^N{\bf U}_{ji}\frac{1}{f_i}{\bf
U}^*_{ik}\right)b_k={\bf a}^*{\bf F^{-1}}{\bf b}
\end{equation}
which proves the statement.
\end{proof}
\begin{prop}
The permanent of a positive definite Hermitian matrix ${\bf F}$
can be represented in terms of a multivariate Gaussian integral as
\begin{equation}\label{permgau1}
Per \,{\bf F}=\frac{1}{\det{\bf F}}\int e^{-{\bf s}^* {\bf
F}^{-1}{\bf s}} \,\prod_{i=1}^N \left(\overline{s}_i s_i\right)\,
d{\bf s}^*d{\bf s}
\end{equation}
\end{prop}
\begin{proof}
According to the contour representation of the permanent
(\ref{main})
 \begin{equation}\label{start1}
 Per \,{\bf F}=\oint e^{{\xi}^*\,{\bf F}\,{\bf
z}}\,\mathcal{D}_N({\bf z},\xi^*).
 \end{equation}
We now replace the exponential factor in the integrand with its
representation according to the Lemma (\ref{lemapp}):
\begin{equation}
e^{{\xi}^*\,{\bf F}\,{\bf z}}=\det{\bf F^{-1}}\,\int e^{-{\bf s}^*
{\bf F}^{-1}{\bf s}-({\bf \xi}^*{\bf s})-({\bf s}^*{\bf z})} d{\bf
s}^*d{\bf s}
\end{equation}
Substituting this back to (\ref{start1}), changing the order of
integrations and straightforwardly performing the contour
integration according to
\begin{equation}
\oint e^{-({\bf \xi}^*{\bf s})-({\bf s}^*{\bf
z})}\,\mathcal{D}_N({\bf z},\xi^*)=\prod_{i=1}^N
\left(\overline{s}_i s_i\right)
\end{equation}
we immediately arrive to (\ref{permgau1}).
\end{proof}
{\bf Note:} The formula (\ref{permgau1}) is in fact a variant of
the Wick theorem for the Gaussian integrals.

Further note that any positive definite Hermitian matrix ${\bf F}$
can be represented as ${\bf F}={\bf E}{\bf E}^*$, where ${\bf
E}={\bf U}{\bf f}^{1/2}$ in terms of the eigenvectors and the
corresponding eigenvalues of ${\bf F}$. This allows us to arrive
at the following result, which seems first to appear in \cite{W}:
\begin{cor}
The permanent of a positive definite Hermitian matrix ${\bf F}$
can be represented as
\begin{equation}\label{bose}
Per\,{\bf F}=\int e^{-{\bf v}^* {\bf v}} \,\prod_{i=1}^N
\left(\left[{\bf E v }^*\right]_i \left[{\bf E v
}\right]_i\right)\, d{\bf v}^*d{\bf v}
\end{equation}
\end{cor}
\begin{proof}
Introduce in (\ref{permgau1}) a new integration variable ${\bf v}$
according to ${\bf s}={\bf E v }$. The integration volume is
changed as $d{\bf s}^*d{\bf s}=\det{\bf E}\det{\bf E}^*d{\bf
v}^*d{\bf v}\equiv \det{F}d{\bf v}^*d{\bf v}$, and the statement
follows.
\end{proof}

\end{document}